\documentclass[aps,superscriptaddress,nofootinbib]{revtex4}
\usepackage{graphicx}	
\usepackage{dcolumn}	
\usepackage{bm}		
\usepackage{amsmath}
\usepackage{amsfonts}
\usepackage{amssymb}
\usepackage{hyperref}
\usepackage{mathtools}
\usepackage{setspace}	
\usepackage{color}
\usepackage{dsfont}
\usepackage{pdfpages}

\definecolor{green}{rgb}{0,.45,0}
\definecolor{orange}{rgb}{1,0.5,0}

\newcommand{\eq}[1]{Eq.~(\ref{#1})}
\newcommand{\be}{\begin{equation}}
\newcommand{\ee}{\end{equation}}
\newcommand{\ba}{\begin{eqnarray}}
\newcommand{\ea}{\end{eqnarray}}
\newcommand{\la}{\langle}
\newcommand{\ra}{\rangle}
\newcommand{\di}{\mathrm{d}}
\newcommand{\Pnew}{P_{\!\Delta c}}

\begin{document}
\title{\boldmath On LHCb pentaquarks as a baryon-$\psi(2S)$ bound state 
-- prediction of isospin $\frac 32$ pentaquarks with hidden charm
\footnote{Devoted to the memory of Prof.\ Alexander Nikolaevich Vall.} }

\author{Irina A.~Perevalova}
	\affiliation{Physics Department, Irkutsk State University, 
		Karl Marx str.~1, 664003, Irkutsk, Russia}
\author{Maxim V.~Polyakov}
	\affiliation{Petersburg Nuclear Physics Institute, 
		Gatchina, 188300, St.~Petersburg, Russia}
	\affiliation{Institut f\"ur Theoretische Physik II, 
		Ruhr-Universit\"at Bochum, D-44780 Bochum, Germany}
\author{Peter~Schweitzer}
	\affiliation{Department of Physics, University of Connecticut, 
		Storrs, CT 06269, USA}
	\affiliation{Institute for Theoretical Physics, T\"ubingen University, 
	Auf der Morgenstelle 14, 72076 T\"ubingen, Germany}

\date{31.\ August 2016}  

%
%
%
%

\begin{abstract} 
The pentaquark $P_c^+(4450)$ recently discovered by LHCb 
has been interpreted as a bound state of $\Psi(2S)$ and nucleon.
The charmonium-nucleon interaction which provides the binding mechanism
is given, in the heavy quark limit, in terms of charmonium chromoelectric 
polarizabilities and densities of the nucleon energy-momentum tensor (EMT). 
In this work we show in model-independent way, by exploring general 
properties of the effective interaction, that  $\Psi(2S)$ can form 
bound states with nucleon and $\Delta$. Using the Skyrme model 
to evaluate the effective interaction in the large-$N_c$ limit and 
estimate $1/N_c$ corrections, we confirm the results from prior work 
which were based on a different effective model (chiral quark soliton model).
This shows that the interpretation of $P_c^+(4450)$ is remarkably 
robust and weakly dependent on the details of the effective theories 
for the nucleon~EMT. We explore the formalism further 
and present robust predictions of isospin $\frac32$ 
bound states of $\Psi(2S)$ and $\Delta$ with masses around $4.5\,{\rm GeV}$ 
and widths around $70\,{\rm MeV}$. The approach also predicts broader
resonances in the $\Psi(2S)$-$\Delta$ channel at $4.9\,{\rm GeV}$ 
with widths of the order of $150\,{\rm MeV}$. We discuss in which reactions
these new isospin $\frac 32$ pentaquarks with hidden charm can be observed.
\end{abstract}

\maketitle


\section{Introduction}
\label{Sec-1:Introduction}

The LHCb collaboration has recently discovered new pentaquark states by 
studying the decays of $\Lambda_b^0 \to J/\Psi\,p\,K^-$ \cite{Aaij:2015tga}.
This decay channel is dominated by the weak decay
$\Lambda_b^0 \to J/\Psi\,\Lambda^\ast$ with subsequent strong decays 
$\Lambda^\ast\to p\,K^-$. 
However, the $J/\Psi\,p$ spectrum contains structures which can be 
interpreted as exotic pentaquark ``$P_c^+$'' $(c\bar c uud)$ resonances.
In about $(8.4\pm0.7\pm4.2)\,\%$ of the cases a broad resonance
$P_c^+(4380)$, and in about $(4.1\pm0.5\pm1.1)\,\%$ 
of the cases a narrow resonance $P_c^+(4450)$ is formed.
Their properties are summarised in Table~\ref{tab:1}.
The analysis of \cite{Aaij:2015tga} is supported by the LHCb 
study \cite{Aaij:2016phn} where it was shown in model-independent way 
that $K^-p$ resonant or nonresonant contributions alone cannot explain 
the stuctures seen in the $\Lambda_b^0 \to J/\Psi\,p\,K^-$ decays.
The recent LHCb analysis of the $\Lambda_b^0 \to J/\psi \, p \, \pi^-$ decays
provides further support for the existence of the new pentaquark states
\cite{Aaij:2016ymb}.

\

\begin{table}[h!]
\begin{center}
{\begin{spacing}{1.7}
\begin{tabular}{|c|c|c|c|c|}
  \hline
  state & mass [MeV] & width [MeV] & isospin & spin-parity $J^P$ \\ \hline
  $P_c^+(4380)$ & $4380\pm8\pm29$ 	& $205\pm18\pm86$ & $\frac12$ 
  & \ \ ${\frac32}^-$ or ${\frac32}^+$ or ${\frac52}^+$ \\
  $P_c^+(4450)$ & $4449.8\pm1.7\pm2.5$ 	& $39\pm5\pm19$   & $\frac12$ 
  & \ \ ${\frac52}^+$ or ${\frac52}^-$ or ${\frac32}^-$ \\ \hline
\end{tabular}\end{spacing}}
\end{center}
\vspace{-7mm}
\label{tab:1}
\caption{Summary of properties of the new pentaquark states observed
at LHCb \cite{Aaij:2015tga}.}
\end{table}

The new states have been interpreted in a variety of theoretical 
approaches. For instance, it was considered that they are loosely bound 
(``molecular'') charmed baryon-meson states \cite{molecula}, bound states 
of light and heavy diquarks including $c$-quarks \cite{diquarks}, and even 
the possibility of open-color bound states was considered \cite{Morozov}.
Also the possibility was discussed that the observed structures could arise 
from threshold cusp effects \cite{cusp}.

In this work we will use the formalism developed in Ref.~\cite{Eides:2015dtr}
where the narrow $P^+_c(4450)$ state was interpreted as a 
nucleon-$\psi(2S)$ $s$-wave bound state with $J^P=\frac32{ }^-$.
In this approach the binding mechanism is provided by the effective 
charmonium-nucleon interaction, which is given by the product of the 
charmonium chromoelectric polarizability and the nucleon EMT densities.
In Ref.~\cite{Eides:2015dtr} also a $J^P=\frac12{ }^-$ state was predicted
with nearly the same mass as $P^+_c(4450)$ 
(modulo hyperfine splitting due to quarkonium-nucleon spin-spin interaction 
which are suppressed in the heavy quark mass limit by $1/m_Q$).
The broader resonance $P_c(4380)$ does not appear as a nucleon-$\psi(2S)$ 
bound state in \cite{Eides:2015dtr}.
Notice that no nucleon-$J/\Psi$ bound states exist in this formalism
as the effective interaction is too weak in this channel.

The purpose of our study is to confirm the findings of 
Ref.~\cite{Eides:2015dtr}, and to investigate whether the formalism
predicts further bound states which could allow us to test this approach. 
For that we will first derive a model-independent 
lower bound which the chromoelectric polarizability must satisfy such 
that charmonium-baryon bound states can~exist. This derivation only 
makes use of general properties of the effective 
baryon-charmonium interaction. We will apply this bound to show 
in model-independent way that $\psi(2S)$ can form $s$-wave bound 
states with nucleon and $\Delta$.

Specific predictions require the use of a model for the non-perturbative
calculation of the EMT densities of baryons. For that in \cite{Eides:2015dtr} 
results were used from the chiral quark soliton model \cite{Goeke:2007fp}. 
In this work, we will use of a different model for EMT densities, 
namely the Skyrme model \cite{Cebulla:2007ei}.
This model is based on chiral symmetry and the $1/N_c$ expansion like 
the chiral quark soliton model. But it differs in many important respects, 
and is therefore well-suited to provide an important cross check.
Our results in the Skyrme model will confirm in detail
the calculation of Ref.~\cite{Eides:2015dtr}.

The chiral soliton model and the Skyrme model describe baryons as 
chiral solitons in the limit of a large number of colors $N_c$, and
provide different practical realizations of the picture of baryons 
in the large-$N_c$ limit of QCD \cite{Witten:1979kh}. In nature $N_c=3$ 
does not seem large, and one may wonder whether $1/N_c$ corrections could 
affect our description of the new pentaquark states. We will therefore
use the Skyrme model to investigate also the role of $1/N_c$ corrections.
For that we will establish a procedure how to construct a conserved EMT 
when a theory cannot be solved exactly and certain (in our case $1/N_c$) 
corrections must be included as a small perturbation. 
We will show that our description of the hidden charm pentaquarks 
is remarkably robust, also when one includes $1/N_c$ corrections.

Our study of the $1/N_c$ corrections to the EMT has interesting by-products.
Soliton models based on the large-$N_c$ expansion describe baryons 
with spin and isospin quantum numbers 
$S=I=\frac12,\;\frac32,\;\frac52,\;\dots\,$ as different rotational 
states of the same soliton solution 
(throughout this work we focus on SU(2) flavor sector).
In contrast to the quantum numbers $S=I=\frac12$ and $\frac32$, 
which correspond respectively to nucleon and $\Delta$,
the quantum numbers $S=I\ge\frac52$ are not observed.
This is considered an unsatisfactory artifact of the (rigid rotator)
soliton approach. Our study will shed new light on this issue.
We will show that $1/N_c$ corrections constitute a 
``reasonably small perturbation'' in the nucleon case. They are
more sizable for $\Delta$ but we find that also in this case
it is possible to reconstruct a conserved EMT which satisfies 
basic criteria for mechanical stability.
However, we will show that for $S=I\ge\frac52$ this is not~possible:
here $1/N_c$ corrections are simply too destabilizing. In this way 
the rotating soliton approach provides an explanation why the quantum 
numbers $S=I\ge\frac52$ are not realized in nature. As another 
by-product we will discuss the EMT of the $\Delta$ and show that 
it has a negative $D$-term in agreement with theoretical studies 
of other particles.

The main application in this context is, however, to investigate the
question whether charmonia can bind with the $\Delta$-resonance. 
We will show that, $J/\Psi$ does not form bound states with $\Delta$. 
But in the $\Delta$-$\psi(2S)$ channel the formalism makes robust
predictions of bound states, and also predicts resonant states 
albeit with somewhat larger theoretical uncertainties. 
We will make specific predictions for the masses, widths and parities 
of the new states.
Finally, we will discuss in which reactions the new states could 
in principle be observed.

\section{Effective quarkonium-baryon interaction}
\label{Sec-2:quarkonium-baryon-iteraction}

In this Section we review the derivation of the effective quarkonium-baryon 
potential and describe how it can be expressed in terms of EMT densities.

\subsection{The effective potential}
\label{Sec-2a:Veff}

The description of hidden-charmonium pentaquark states of 
Ref.~\cite{Eides:2015dtr} 
explores the fact that heavy charmonium states are small compared to the 
nucleon size, and their interaction with baryons is relatively weak on the 
typical scale for strong interactions. In this situation a non-relativistic 
multipole expansion can be applied \cite{Gottfried:1977gp}. 

The multipole expansion reveals that the dominant mechanism for the 
baryon-quarkonium interaction is the emission of two virtual chromoelectric 
dipole gluons in a color singlet state. The potential describing the 
effective interaction is proportional to the product of the quarkonium 
chromoelectric polarizability and the gluon energy-momentum density 
in the nucleon \cite{Voloshin:2007dx}. The small parameter justifying
this derivation is given by the ratio of the quarkonium size to the 
effective gluon wave-length. The resulting effective dipole
Lagrangian is given by \cite{Voloshin:1979uv}
\be\label{Eq:Leff-1}
	L_{\rm eff} = -\,V_{\rm eff} \, , \;\;\;
	V_{\rm eff} = -\,\frac12\;\alpha\;\bm{E}^2
\ee
where $\alpha$ denotes the chromoelectric polarizability in the channel
of interest, and $\bm{E}$ is the chromoelectric gluon field, whose 
definition includes the strong coupling constant $g$ renormalized 
at the quarkonium mass scale. 

\subsection{Chromoelectric polarizabilities}
\label{Sec-2b:polarizabilities}

The chromoelectric polarizabilities can be calculated in the heavy 
quark approximation and large-$N_c$ limit, where the quarkonia are 
described as Coulomb systems in lowest order approximation, with the 
results given by \cite{Peskin:1979va,Eides:2015dtr}
\begin{subequations}
\label{Eq:alpha-guideline}
\begin{align}
	\label{Eq:alpha-1S}
	\alpha(1S) & \approx    0.2\,{\rm GeV}^{-3}\;\mbox{(pert)}, \\
	\label{Eq:alpha-2S}
	\alpha(2S) & \approx \; 12\, {\rm GeV}^{-3}\;\mbox{(pert)}, \\
	\label{Eq:alpha-2S-to-1S}
	\alpha(2S\to 1S) & \approx 
		\begin{cases}
		 -0.6\, {\rm GeV}^{-3}\;\mbox{(pert)},\\
		 \pm 2\,{\rm GeV}^{-3}\;\mbox{(pheno)}.
		\end{cases} 
\end{align}
\end{subequations}
In Eq.~(\ref{Eq:alpha-2S-to-1S}) we included also the phenomenological 
value for the polarizability of the $2S \to 1S$ transition inferred from 
analyses of $\psi^\prime\to J/\psi \pi\pi$ data \cite{Voloshin:2007dx}
(such studies only allow to extract the modulus of the
transitional polarizability). 
For $\alpha(1S)$ the $1/N_c$ corrections are merely of
order of $5\,\%$ \cite{Brambilla:2015rqa}. But for $\alpha(2S)$ 
and higher polarizabilities the effects of $1/N_c$ corrections
are not known, and the comparison of the perturbative and
phenomenological results in Eq.~(\ref{Eq:alpha-2S-to-1S}) indicates that 
at present the chromoelectric polarizabilities are not well understood. 
Below we will therefore use the values quoted in 
Eq.~(\ref{Eq:alpha-guideline}) not at their bare values, but as guidelines. 

For $\psi(nS)$  with $n\ge 3$ the perturbative results for 
polarizabilities grow rapidly with $n$ as $\alpha(nS)\propto n^2(7n^2-3)$
\cite{Peskin:1979va} because the size of the system grows.
In this situation the Coulomb approximation becomes worse, and the
usefulness of perturbative predictions for $\psi(3S)$ and higher 
states becomes questionable.

\subsection{Relation to EMT densities}
\label{Sec-2c:Veff+densities}

The effective interaction in Eq.~(\ref{Eq:Leff-1}) can be expressed 
in terms of the densities of the nucleon EMT.
This can be done exploring the conformal anomaly \cite{Nielsen:1977sy} 
to relate $\bm{E}^2$ in Eq.~(\ref{Eq:Leff-1}) to the trace ${T^\mu}_\mu$
of the EMT of QCD and the gluon contribution to the energy density $T_{00}^G$. 
The latter can be related as $T^G_{00} = \xi_s\;T_{00}$ to the total energy 
density $T_{00}$ of the nucleon where the parameter $\xi_s$ describes the 
fraction of the nucleon energy due to gluons at the scale $\mu_s$
\cite{Novikov:1980fa}.
Neglecting a numerically small term due to the current masses of
light quarks one obtains \cite{Eides:2015dtr},
\be\label{Eq:expressing-E}
	\bm{E}^2 
	= g^2\biggl(\frac{8\pi^2}{b\,g_s^2}\,{T^\mu}_\mu + \xi_s\;T_{00}\biggr) 
	= \frac{8\pi^2}{b}\,\frac{g^2}{g_s^2}\biggl(\nu\,T_{00}+{T^k}_k\biggr)
	\, , \;\;\; \nu = 1+\xi_s\,\frac{b\,g_s^2}{8\pi^2} \; ,
\ee
where $b = (\frac{11}{3}-\frac{2}{3}\,N_f)$ is the leading coefficient 
of the Gell-Mann-Low function, and $g_s$ is the strong coupling constant 
renormalized at the scale $\mu_s$. Notice that the relevant scale for 
non-perturbative calculations of the nucleon structure $\mu_s$
is different from the quarkonium scale at which the strong coupling $g$ 
is renormalized. Recall that $g$ enters Eq.~(\ref{Eq:expressing-E}) 
through the definition of the chromoelectric gluon field $\bm{E}$. 
Therefore in general $g_s\neq g$ although for the charmonium-nucleon 
potential these 2 scales are comparable.

The coefficient $\nu$ introduced in Eq.~(\ref{Eq:expressing-E}) was
estimated on the basis of the instanton liquid model of the QCD 
vacuum and the chiral quark soliton model, where the strong coupling 
constant freezes at scale set by the nucleon size at 
$\alpha_s = g_s^2/(4\pi) \approx 0.5$. Assuming $\xi_s\approx 0.5$ as 
suggested by the fraction of nucleon momentum carried by gluons in DIS 
at scales comparable to $\mu_s$ one obtains the value \cite{Eides:2015dtr} 
\be\label{Eq:nu}
	\nu \approx 1.5 \,.
\ee
A similar result $\nu = $(1.45--1.6) was obtained for the pion in 
Ref.~\cite{Novikov:1980fa}. These results are supported by the
analysis of the nucleon mass decomposition in Ref.~\cite{Ji:1994av} 
where $\xi_s\approx \frac13$ leading to $\nu \approx 1.4$ 
which is within the accuracy of Eq.~(\ref{Eq:nu}).
We will use the value (\ref{Eq:nu}) for the calculations in this work.

\newpage
\section{Sufficient condition for existence of a quarkonium-baryon bound state}
\label{Sec-3:condition-for-bound-state}

In this Section we discuss,  in a model-independent way, the lower bound for 
the chromoelectric polarizability at which a quarkonium-baryon bound state 
is formed.  In Ref.~\cite{Calogero}  the following sufficient condition 
for the existence of an $s$-wave bound state in a given attractive potential 
was derived
\begin{equation}\label{Col}
  	-\frac{2\mu}{R}\int\limits_{0}^{R}\mathrm{d}r\, r^2 V(r)
	-2\mu R \int\limits_{R}^{\infty}\mathrm{d}r\, V(r)> 1\, ,
\end{equation}
where $R$ is an arbitrary distance, $\mu$ is the reduced baryon-charmonium 
mass, and the attractive potential $V(r)$ is negative. We will refer to
this condition as Calogero bound in the following.

Let us consider first the nucleon case. The effective 
$\psi(2S)$-nucleon potential is normalized as \cite{Eides:2015dtr} 
\begin{equation}\label{norm}
	\int\limits_{0}^{\infty}\mathrm{d}r\, r^2 V_{\rm eff}(r) =
	-\alpha\frac{\pi}{b} \frac{g^2}{g_s^2} \nu M_N \, ,
\end{equation}
also its large $r$ asymptotics (in the chiral limit
in leading order of the large-$N_c$ expansion) is known 
\cite{Eides:2015dtr} (see also Eq.~(\ref{Eq:Veff-large-r}) below):
\begin{equation}\label{asy}
 	V_{\rm eff}(r)\sim -\alpha\frac{{27}}{4\, b} \frac{g^2}{g_s^2}(1+\nu)
	\frac{g_{A}^{2}}{F_{\pi}^2 \, r^6}\, .
\end{equation}
Now we can choose the parameter $R$ in Eq.(\ref{Col}) large enough such 
that for $r>R$ the asymptotics (\ref{asy}) can be used. This allows us to 
rewrite Eq.(\ref{Col}) as an inequality for the chromoelectric polarizability:
\begin{equation}\label{ineq}
 	\alpha>\frac{b}{2\pi} \frac{g_s^2}{g^2}\frac{1}{\nu}\frac{R}{\mu M_N}
	{\left[1-\frac{9}{10\pi}\frac{1+\nu}{\nu}
	\frac{g_{A}^2}{F_{\pi}^2 M_N}\frac{1}{R^3}\right]^{-1}}\, .
\end{equation}
Note that this inequality is model independent as it is based only on 
general (model independent) properties of the effective potential 
(\ref{norm}) and (\ref{asy}).
If we choose $R=1.5$~fm (for that value we are sure  that asymptotic 
formula (\ref{asy}) works perfectly) 
and take the non-commutativity of the chiral limit and the 
large-$N_c$ limit (see App.~\ref{App-A:large-r}) into consideration, 
we obtain that for 
$\alpha> 10.7 $~GeV$^{-3}$ the nucleon and $\psi(2S)$ must form a bound state.  
This value for the lower bound does not depend on details 
of the potential shape. 

The inequality (\ref{ineq}) can be easily generalized to any other baryon. 
What one needs for that is to derive the large distance behavior of 
$V_{\rm eff}(r)$ for a given baryon.\footnote{
	For a baryon of mass $M_B$ the normalization condition is trivial 
$$
	\int\limits_{0}^{\infty}\mathrm{d}r\, r^2 V_{\rm eff}(r)
	=-\alpha \frac{\pi}{b} \frac{g^2}{g_s^2}\nu M_B\, .
$$}
This can be done with help of Chiral Perturbation Theory. 

For example, if one applies the Calogero lower bound to the case of 
the $\Delta$-resonance, one obtains that  for $\alpha>6.6$~GeV$^{-3}$
a charmonium-$\Delta$ bound state must form, i.e. the formation of such 
bound state with isospin 3/2 is more favourable than for the nucleon.

\newpage
\section{Energy-momentum tensor and EMT densities}
\label{Sec-4:EMT-in-general}

In this section we briefly introduce the form-factors of the EMT,
define the static EMT and the EMT densities, and review their
properties which are relevant for our study.

\subsection{Form factors and EMT densities}
\label{Sec-3a:FFs+densities}

The nucleon form factors of the total EMT operator
$\hat T_{\mu\nu}(0)$ are defined as \cite{Pagels}
\be\label{Eq:ff-of-EMT} 
    	\la p^\prime,s^\prime| \hat T_{\mu\nu}(0) |p,s\rangle
    	= \bar u(p^\prime,s^\prime)\biggl[M_2(t)\,\frac{P_\mu P_\nu}{M_N} + J(t)\,
	\frac{i(P_{\mu}\sigma_{\nu\rho}+P_{\nu}\sigma_{\mu\rho})\Delta^\rho}{2M_N}
	+ d_1(t)\,\frac{\Delta_\mu\Delta_\nu-g_{\mu\nu}\Delta^2}{5M_N}\biggr]u(p,s)
	\, , \ee
where $P=\frac12(p'+p)$, $\Delta=(p'-p)$, $t=\Delta^2$ 
with nucleon states normalized as $\la p^\prime,s^\prime|p,s\ra =  
2p^0(2\pi)^3\delta^{(3)}(\bm{p}^\prime-\bm{p}) \delta_{s^\prime s^{ }}$. 
The polarizations $s$ and $s^\prime$ are defined such that both 
correspond to the same polarization vector $\bm{s}$ in the rest 
frame of the corresponding nucleon. The spinors are normalized
as $\bar u(p,s) u(p,s)=2 M_N$.

In QCD the quark and gluon contributions to the EMT are separately 
gauge-invariant operators and connected to observables, although only 
their sum is scale-independent and conserved. They can be deduced from
Mellin moments of the generalized parton distribution functions of quarks
and gluons accessible in hard exclusive reactions.

In analogy to the electromagnetic form factors one may introduce the 
static EMT in the Breit frame characterized by $\Delta^0=0$ which implies
$t=-\bm{\Delta}^2$. In this frame the static EMT is defined\footnote{
	Notice the misprint in Eq.~(5) of \cite{Polyakov:2002yz} where
	the factor $1/(2E)$ should appear under the integral, as 
	written in Eq.~(\ref{Eq:static-EMT}).} 
as \cite{Polyakov:2002yz}
\be\label{Eq:static-EMT}
	T_{\mu\nu}(\bm{r},\bm{s}) = \int \frac{\di^3\bm{\Delta}}{2E(2\pi)^3}
	\,e^{i\bm{\Delta}\bm{r}}\la p^\prime,s^\prime| \hat T_{\mu\nu}(0) |p,s\rangle
\ee
where $E=E^\prime = \sqrt{M_N^2+\frac14\bm{\Delta}^2}$. Working with
3-dimensional densities, which strictly speaking are well-defined
only for non-relativistic systems, is fully consistent in our context
because we will use models for the EMT based on the large-$N_c$ limit, 
where baryons are heavy. 
This is also fully consistent with the non-relativistic interaction 
(\ref{Eq:Leff-1}) of heavy quarkonia with baryons, and the guidelines 
(\ref{Eq:alpha-guideline}) for polarizabilities calculated for heavy 
quarkonia in large-$N_c$ limit.

Let us review here the densities relevant for this work, namely
the energy density $T_{00}(\bm{r})$ and the stress tensor $T^{ij}(\bm{r})$.
For a more detailed discussion of the static EMT we refer to 
\cite{Polyakov:2002yz}. The energy density is normalized as
\be\label{Eq:T00-mass}
	\int\di^3\bm{r}\;T_{00}(\bm{r}) = M_N \,.
\ee
For a spin $\frac12$ particle (as well as for a spin 0 particle) the
stress tensor has the general decomposition 
\be\label{Eq:Tij}
	T^{ij}(\bm{r}) = \left(e_r^ie_r^j-\frac13\,\delta^{ij}\right)
	s(r) + \delta^{ij}\,p(r)\,
\ee
where $p(r)$ is the pressure and $s(r)$ is the distribution of shear forces,
while $e_r^i=r^i/r$ denotes the radial unit vector and $r=|\bm{r}|$.

\subsection{Consequences from EMT conservation}
\label{Sec-3b:EMT-conservation}

Due to EMT conservation $s(r)$ and $p(r)$ are related to each other through 
the differential equation
\be\label{Eq:diff-eq-s-p}
 	\frac{2}{r}\,s(r) + \frac{2}{3}\,s^\prime(r) + p^\prime(r) = 0\;,
\ee
and $p(r)$ obeys \cite{Goeke:2007fp} the von Laue condition \cite{von-Laue},
a necessary (though not sufficient) condition for stability,
\be\label{Eq:von-Laue}
      \int_0^\infty \di r\;r^2p(r)=0\;.
\ee
In order to comply with (\ref{Eq:von-Laue}) the pressure must 
have at least one node. Stability considerations imply that 
$p(r)>0$ in the inner region which corresponds to repulsion, 
and $p(r)<0$ in the outer region which corresponds to attraction,
with the repulsive and attractive forces balancing exactly 
according to Eq.~(\ref{Eq:von-Laue}) \cite{Goeke:2007fp}.
An interesting quantity related to the stress tensor is the $D$-term,
which is a fundamental but unknown property \cite{Polyakov:1999gs} and 
expressed in terms of the pressure {\it or} shear forces as 
\cite{Polyakov:2002yz}
\ba
    d_1 &=& \phantom{-}\; 5\pi\,M_N
	\int_0^\infty\di r\;r^4\,p(r) \label{Eq:def-d1-pressure}\; \\
    	&=& -\,\frac{4\pi}{3}\,M_N
	\int_0^\infty\di r\;r^4\,s(r) \label{Eq:def-d1-shear}\;.
\ea 
In all theoretical approaches so far the $D$-terms of various
particles were found negative.

In all expressions presented so far $s(r)$ and $p(r)$ appear on equal 
footing. As long as one deals with the total EMT in a consistently solved
theory, both quantities are indeed related to each other and completely
equivalent. However, in some situations one may deal with an incomplete
system. One example is when one considers form factors of the quark-part
of the EMT in QCD. Another situation may arise when one is not able to
find the exact solution but has to content oneself with an approximate
solution in a (effective) theory. (We will encounter exactly this
situation~below.) 

In such situations, working with $s(r)$ is preferable
over $p(r)$ for the following reason. If one deals with only a part
of the system, e.g.\ with the quark contribution to the EMT, then
there is a fourth form factor in Eq.~(\ref{Eq:ff-of-EMT}) which
is proportional to the structure $g^{\mu\nu}$ (the gluon part 
of the EMT has the same form factor but with opposite sign, such 
that in the total quark + gluon EMT these terms drop out).
Now we have seen that the pressure is associated with the trace 
of the stress tensor, and is sensitive to terms
arising from non-conservation of the EMT. In contrast to this,
$s(r)$ is associated with the traceless part of the stress tensor,
and is therefore insensitive to EMT-nonconserving terms.
Below we will use this property to {\it reconstruct} from 
approximate results for $s(r)$ a conserved EMT.

\subsection{Local criteria for stability}
\label{Sec-3c:local-criteria}

When constructing effective theories or models it is essential to demonstrate 
their theoretical consistency. Hereby the perhaps most important point 
concerns the stability of the studied solution. The von Laue condition 
(\ref{Eq:von-Laue}) provides a useful global criterion, which was shown to 
be satisfied in various approaches including nuclei, nucleons, pions, 
Skyrmions, $Q$-balls where the solutions were absolutely stable 
\cite{Goeke:2007fp,Goeke:2007fq,Cebulla:2007ei,Jung:2013bya,Kim:2012ts,
Jung:2014jja,Mai:2012yc}.
But also meta-stable and unstable solutions satisfy the von Laue 
condition \cite{Mai:2012yc,Mai:2012cx,Cantara:2015sna} which means it
is a necessary but not sufficient condition for stability.

For our purposes it will be convenient to establish a necessary
local stability condition. Local in our context means that it is
not an integrated over $r$ like the von Laue condition.
For that we explore the analogy to classical continuum theory. 
This is well-justified in our context, since we have in mind to apply 
the criteria to a semi-classical description of the nucleon in terms 
of a large-$N_c$ mean field solution. 
An intuitive criterion is the positivity of the energy density 
\be\label{Eq:local-criterion-0}
	T_{00}(r) \ge 0 \,.
\ee
In classical continuum mechanics it follows from considering 
that every (also infinitesimally small) piece of volume
makes a positive contribution to the energy of the system.

A less trivial local criterion can be obtained by considering that at any 
chosen distance $r$ the force exhibited by the system on an infinitesimal 
piece of area $\di A \,e_r^i$ must be directed outwards. If this was not 
the case, the system would collapse. 
Since this force is 
$F^i(\bm{r})=T^{ij}(\bm{r})\di A \,e_r^j=[\frac23\,s(r)+p(r)]\di A\,e_r^i$
we obtain the criterion
\be\label{Eq:local-criterion-1}
	\frac{2}{3}\,s(r) + p(r) > 0 \, .
\ee
We checked that the condition (\ref{Eq:local-criterion-1}) is satisfied 
in all systems we are aware of where EMT densities were studied 
\cite{Goeke:2007fp,Goeke:2007fq,Cebulla:2007ei,Jung:2013bya,
Kim:2012ts,Jung:2014jja,Mai:2012yc,Mai:2012cx,Cantara:2015sna}.
As this includes unstable systems, apparently also (\ref{Eq:local-criterion-1})
is a necessary but not sufficient condition for stability. Due to
its local character, it provides a stronger criterion than the von 
Laue condition (\ref{Eq:von-Laue}) and will play an important role 
below. Interestingly, the criterion (\ref{Eq:local-criterion-1}) 
allows one to draw a conclusion on the sign of the $D$-term. 
We see that 
\be\label{Eq:sign-d1}
	0 < 4\pi \int_0^\infty\di r\;r^4\biggl(\frac{2}{3}\,s(r) + p(r)\biggr) 
	= -\frac{2d_1}{M_N}+\frac{4d_1}{5M_N} = -\,\frac{6d_1}{5M_N} \;.
\ee
Thus, if a system satisfies the local stability criterion 
(\ref{Eq:local-criterion-1}), then it must necessarily have a negative $D$-term
(but a negative $D$-term does not imply that $s(r)$ and $p(r)$ satisfy 
(\ref{Eq:local-criterion-1}), so the opposite is in general not true).
Indeed, in all systems studied so far the $D$-terms were found to be
negative \cite{Goeke:2007fp,Goeke:2007fq,Cebulla:2007ei,Jung:2013bya,Kim:2012ts,
Jung:2014jja,Mai:2012yc,Mai:2012cx,Cantara:2015sna}.

It would be natural to expect that the criteria 
(\ref{Eq:local-criterion-0},~\ref{Eq:local-criterion-1}) hold also in 
quantum field theory, although in this case more care is needed.
Investigations in this direction are left to future studies.

\subsection{Chiral properties of densities}
\label{Sec-3d:chiral-properties}

The leading large-distance dependence of the densities is determined 
by chiral physics, and can be derived in any (effective) theory which
consistently describes chiral symmetry breaking. Soliton models are 
particularly convenient for that \cite{Goeke:2007fp,Cebulla:2007ei}.
In the chiral limit in leading order of the large-$N_c$ expansion
the densities behave as, see App.~\ref{App-A:large-r},
\begin{subequations}\label{Eq:EM-FF-large-r-0}
\begin{align}
	\label{Eq:T00-large-r-0}
	T_{00}(r)& = 3\;F_\pi^2\;R_0^4\;\frac{1}{r^6} + \dots\; \,,\\
	\label{Eq:p-large-r-0}
	p(r) 	& = -\,F_\pi^2\;R_0^4\;\frac{1}{r^6} + \dots\;\,,\\
	\label{Eq:s-large-r-0}
	s(r) 	& = 3\;F_\pi^2\;R_0^4\;\frac{1}{r^6} + \dots\; \,,
\end{align}
\end{subequations}
where the dots indicate terms vanishing faster than the displayed
leading terms. The parameter $R_0$ has the meaning of the soliton
size in chiral soliton models, and is related to the axial coupling 
constant $g_A=1.26$ and the pion decay constant $F_\pi=186\,$MeV as 
\be\label{Eq:R0-gA}
	g_A = \frac{4\pi}{3} \, F_\pi^2 \, R_0^2 \;.
\ee
In practice one has to determine $R_0$ from the self-consistent
profile which minimizes the soliton energy (we will discuss this 
in more detail in Sec.~\ref{Sec-5a:Skyrme-with-Nc-corrections}), and 
Eq.~(\ref{Eq:R0-gA}) can be used to deduce the model prediction for $g_A$.
For finite $m_\pi$ the densities exhibit exponentially suppressed Yukawa
tails, see App.~\ref{App-A:large-r}.

\subsection{\boldmath $V_{\rm eff}$ and its properties}
\label{Sec-3e:Veff-properties}

We are now in the position to express the effective potential $V_{\rm eff}$ 
in Eq.~(\ref{Eq:Leff-1}) in terms of the EMT densities. With the trace of 
the stress tensor given by ${T^k}_k=-\,3\,p(r)$ the effective potential is
\be\label{Eq:Veff}
	V_{\rm eff}(\bm{r}) = -\,\alpha\;\frac{4\pi^2}{b}\,\frac{g^2}{g_s^2}
	\biggl(\nu\,T_{00}(r)-3\,p(r)\biggr)  .
\ee
Due to Eqs.~(\ref{Eq:T00-mass},~\ref{Eq:von-Laue}) the effective
potential is ``normalized'' as 
\be\label{Eq:Veff-norm}
	\int\di^3r\;V_{\rm eff}(\bm{r}) = -\,\alpha\;
	\frac{4\pi^2}{b}\,\frac{g^2}{g_s^2}\;\nu\,M_N \, .
\ee
An instructive property of the effective potential, which may provide a
 useful estimate for the ``range'' of the effective interaction, is the 
mean square radius
\be\label{Eq:Veff-r2}
	\la r^2_{\rm eff}\ra \equiv 
	\frac{\int\di^3r\;r^2V_{\rm eff}(\bm{r})}{\int\di^3r\;V_{\rm eff}(\bm{r})}
	= \la r_E^2\ra - \frac{12\, d_1}{5\nu M_N^2}
\ee
where $\la r_E^2\ra = \int\di^3r\;r^2T_{00}(r) / \int\di^3r\;T_{00}(r)$ 
denotes the mean square radius of the energy 
density. With $d_1<0$ found so far in all theoretical studies, 
one may expect $\la r^2_{\rm eff}\ra > \la r^2_E\ra$.

From Eqs.~(\ref{Eq:T00-large-r-0},~\ref{Eq:p-large-r-0}) we see that
in the chiral limit the effective potential behaves as
\be\label{Eq:Veff-large-r}
	V_{\rm eff}(\bm{r}) = -\,\alpha\;\frac{12\pi^2}{b}\,\frac{g^2}{g_s^2}\;
	(1+\nu) \;F_\pi^2\;R_0^4\;\frac{1}{r^6} \; + \; \dots\;\; .
\ee
Using (\ref{Eq:R0-gA}) one obtains Eq.~(\ref{asy}) 
quoted in Sec.~\ref{Sec-3:condition-for-bound-state}.

\section{\boldmath EMT of nucleon and $\Delta$ in Skyrme model}
\label{Sec-5:Skyrme}

In order to solidify the predictions from \cite{Eides:2015dtr} 
and gain new insights on the baryon-charmonium interaction, we will 
use the Skyrme model \cite{Skyrme:1961vq}, which respects chiral symmetry
and provides a practical realization of the large-$N_c$ picture of baryons 
described as solitons of mesonic fields \cite{Witten:1979kh}.
Despite its long history dating back to
\cite{Adkins:1983ya,Adkins:1983hy,Guadagnini:1983uv,
Adkins:1984cf,Bander:1984gr,Braaten:1984qe,Zahed:1986qz}
this model still provides good services, and was applied to studies of 
the EMT in \cite{Cebulla:2007ei} which we shall explore in this work.

\subsection{Description of baryons in Skyrme model}
\label{Sec-5a:Skyrme-with-Nc-corrections}

In this Section we briefly review the description of baryons 
in the Skyrme model. For a detailed account we refer to 
\cite{Adkins:1983ya,Adkins:1983hy}.
The Skyrme model is based on the following effective chiral Lagrangian 
\be\label{Eq:Lagrangian}
    {\mathcal L}=
        \frac{F_\pi^2}{16}\;{\rm tr_F}(\partial_\mu U)(\partial^\mu U^\dag)
    +  \frac{1}{32e^2} \;{\rm tr_F}
    [U^\dag(\partial_\mu U),U^\dag(\partial_\nu U)]\,
    [U^\dag(\partial^\mu U),U^\dag(\partial^\nu U)]
    +  \frac{m_\pi^2F_\pi^2}{8}\;{\rm tr_F}(U-2) \, .
\ee
Here $F_\pi$ the pion decay constant whose experimental value is
$F_\pi=186\,{\rm MeV}$, $e$ is the dimensionless Skyrme parameter, 
$m_\pi$ is the pion mass, and ${\rm tr_F}$ denotes the trace over 
SU(2) matrices. In the large-$N_c$ limit the model parameters
scale as $F_\pi = {\mathcal O}(N_c^{1/2})$, $m_\pi = {\mathcal O}(N_c^0)$, 
$e = {\mathcal O}(N_c^{-1/2})$ which implies ${\mathcal L}={\mathcal O}(N_c)$.
In the large-$N_c$ limit the chiral SU(2) field $U$ is static, and assumed 
to have the ``hedgehog'' structure $U=\exp[i\bm{\tau}\bm{e_r} P(r)]$ 
with $r=|\bm{r}|$ and $\bm{e_r}=\bm{r}/r$. The soliton profile $P(r)$
satisfies $P(0)=\pi$ which ensures that the field $U$ has unit 
winding number associated with the baryon number.
The large distance behavior of $P(r)$ is dictated by chiral symmetry 
and model-independent, see App.~\ref{App-A:large-r}.

In leading order of the large-$N_c$ limit the soliton mass is given by
$M_{\rm sol}=-\int\di^3r\;{\mathcal L} \equiv \int\di^3r\,T_{00}(r)$
and the variation of the soliton mass, $\delta M_{\rm sol}=0$, is
exactly equivalent to the von Laue condition (\ref{Eq:von-Laue}).
This was proven analytically and confirmed numerically in 
Ref.~\cite{Cebulla:2007ei} where the expressions for $T_{00}(r)$, $p(r)$ 
and other EMT densities were derived  and evaluated in leading (LO)
and next-to-leading order (NLO) of the large-$N_c$ expansion.

The minimization of the soliton mass $\delta M_{\rm sol}=0$ yields
the soliton solution which is then projected on spin and isospin 
quantum numbers by considering slow rotations 
$U(\bm{r}) \to A(t)U(\bm{r})A^{-1}(t)$ in (\ref{Eq:Lagrangian})
with $A = a_0+i\,\bm{a}${\boldmath $\tau$}, and introducing conjugate 
momenta $\pi_b=\partial L/\partial\dot{a}_b$. One then quantizes the 
collective coordinates according to $\pi_b \to -i \partial /\partial a_b$ 
subject to the constraint $a_0^2+\bm{a}^2=1$. This yields the Hamiltonian
for soliton rotations
\be\label{Eq:H-rot}
    H_{\rm rot} = M_{\rm sol} + \frac{\bm{J}^2}{2\Theta}
              = M_{\rm sol} + \frac{\bm{I}^2}{2\Theta}
\ee
where $\bm{J}^2$ and $\bm{I}^2$ are the squared spin and isospin operators,
and $\Theta$ denotes the soliton moment of inertia which is a functional 
of the soliton profile.
The Hamiltonian (\ref{Eq:H-rot}) describes states with the spin $S$ 
and isospin $I$ quantum numbers $S=I=\frac12,\,\frac32,\,\dots\;$
with the highest possible spin equal to $N_c/2$ for general $N_c$.
Clearly, isospin quantum numbers $I>\frac32$ are exotic and correspond
to hypothetical multipletts that are not observed in nature. We shall
come back to this point below.

The above described procedure corresponds to the ``projection after variation''
technique used in most practical applications. Indeed, Eq.~(\ref{Eq:H-rot})
 implies that the mass of a baryon with quantum numbers 
$S=I=\frac12,\,\frac32,\,\dots$ is given by
\be\label{Eq:Mrot}
	M_{\rm rot}=M_{\rm sol}+\frac{S(S+1)}{2\Theta}
\ee
with the ``correction'' due to soliton rotations assumed to be a 
small perturbation. Parametrically this is the case, since the moment of 
inertia is $\Theta={\cal O}(N_c)$ and we work in the large-$N_c$ limit.
But in practice it is $N_c=3$ and the ``perturbation'' is not
necessarily small in all cases. A particularly sensitive quantity
in this respect is the pressure. Including systematically $1/N_c$
corrections to the EMT modifies not only $T_{00}(r)$ leading to 
Eq.~(\ref{Eq:Mrot}), but also $p(r)$ and $s(r)$.
The pressure with included $1/N_c$ corrections satisfies the von Laue
condition~(\ref{Eq:von-Laue}) only if one minimizes the full expression
in Eq.~(\ref{Eq:Mrot}). However, in the projection-after-variation
technique one only minimizes $M_{\rm sol}$, and ``rotational corrections'' 
strictly speaking spoil stability \cite{Cebulla:2007ei}.

In principle, one could use the ``variation after projection''-technique
to remedy this problem. Here one minimizes the mass of the rotating
baryon in Eq.~(\ref{Eq:Mrot}), i.e.\ performs first the projection 
on the quantum numbers of the considered baryon before minimizing its 
mass. In this way one ensures compliance with the von Laue condition 
(\ref{Eq:von-Laue}).
However, this procedure has a serious drawback: it is at variance with 
chiral symmetry as can be seen from the large-$r$ behavior of the profile 
\cite{Bander:1984gr,Braaten:1984qe} 
\be\label{Eq:P(r)-at-large-r}
	F(r) = \frac{\rm const}{r}\,\exp(-m_S r)\;\;\;{\rm with}\;\;\;
	m_S^2 = m_\pi^2 - \frac{2S(S+1)}{3 \,\Theta[P(r)]^2} \; .
\ee
Since $\Theta={\cal O}(N_c)$ we see that for $N_c\to\infty$ 
we have $m_S\to m_\pi$ and recover from Eq.~(\ref{Eq:P(r)-at-large-r}) 
the correct chiral behavior of the profile, see App.~\ref{App-A:large-r}.
But for finite $N_c$ the result is incorrect, and for small $m_\pi$ 
solutions do not even exists.

Chiral symmetry and stability are crucial principles.
If one wants to preserve both, then none of the 2 methods, 
``projection after variation'' and ``variation after projection,'' 
is acceptable. In our context, however, we are mainly interested
in gaining trustful insights on effects of $1/N_c$ corrections 
on the effective baryon-quarkonium interaction.
For that reason, we will content ourselves with a pragmatic 
approximate solution, which (a) preserves chiral symmetry, (b) 
complies with the von Laue condition, and (c) gives us reliable 
insights on the role of $1/N_c$ corrections.

The approximate solution, which fullfils the above criteria, is as 
follows. In the first step we employ the ``projection after variation'' 
procedure which respects chiral symmetry. This means we first minimize 
the soliton energy, which guarantees the correct chiral behavior of the 
theory and yields a {\it universal profile for all (light) baryons}. 
After this we project the soliton solution on 
$S=I=\frac12,\,\frac32\,\dots$ states 
(nucleon, $\Delta$-resonance, $\dots$ where the dots indicate
exotic quantum numbers not observed in nature).
This yields, in the leading order of the large-$N_c$ limit,
the same EMT for all (light) baryons. 

In the second step we then add on top of the leading order 
results the rotational corrections as a {\it small perturbation}.
We do so for the energy density $T_{00}(r)$ and shear forces $s(r)$, 
but not for the pressure because the resulting $p(r)$ would 
violate the von Laue condition (\ref{Eq:von-Laue}). Instead, 
we determine the pressure from the differential equation 
(\ref{Eq:diff-eq-s-p}), which then automatically satisfies 
the von Laue condition (\ref{Eq:von-Laue}).
Using (\ref{Eq:diff-eq-s-p}) the pressure can be expressed in terms 
of $s(r)$ as 
(notice that (\ref{Eq:diff-eq-s-p}) determines $p(r)$ up to an integration 
constant, which we fix such that $p(r)\to 0$ for $r\to\infty$)
\be\label{Eq:p-reconstructed}
	p(r)\biggl|_{\rm NLO,reconstruct} 
	= \left(-\,\frac23\,s(r) + 2\int\limits_r^\infty
	\frac{\di \tilde{r}}{\tilde{r}}\,s(\tilde{r})\right)_{\rm NLO, approx}\,.
\ee
This procedure corresponds to the construction of a conserved EMT 
from approximate results for $s(r)$. It is important to notice that the
starting point for this construction is $s(r)$, which is related to the
traceless part of the stress tensor and therefore in general insensitive 
to EMT-nonconserving terms, see Sec.~\ref{Sec-3b:EMT-conservation}.
Below we will show that this procedure gives a consistent 
estimate of $1/N_c$ corrections to EMT densities.

\subsection{\boldmath EMT densities with $1/N_c$ corrections}

The expressions for the EMT densities in leading (LO) and subleading (NLO)
order of the $1/N_c$ expansion were derived in \cite{Cebulla:2007ei}.
We refer to this work for technical details and use the same 
parameters\footnote{\label{Footnote:parameters} 
	The parameters are fixed as $F_\pi = 131.3\,{\rm MeV}$, $e = 4.628$
	with $m_\pi=138\,{\rm MeV}$. This parameter choice has been optimized 
	\cite{Cebulla:2007ei} to ensure that the respective leading results
	in the large-$N_c$ expansion for the sum and difference of nucleon- 
	and $\Delta$-masses, namely $M_\Delta+M_N\equiv 2M_{\rm sol}$ and 
	$M_\Delta-M_N\equiv\frac{3}{2\Theta}$,
	reproduce the experimental values. Notice that with this fixing
	the experimental value of $F_\pi=186\,{\rm MeV}$ is underestimated
	by $30\,\%$ while the model results for the individual nucleon- and 
	$\Delta$-masses overestimate the physical values by about $20\,\%$
	\cite{Cebulla:2007ei}. This is a typical accuracy for this model
	\cite{Adkins:1983ya,Adkins:1983hy}. The $20\,\%$-overestimate of 
	the baryon masses would affect the normalization of the effective
	potential $V_{\rm eff}$ in Eq.~(\ref{Eq:Veff-norm}), which we shall
	address below in Eq.~(\ref{Eq:Veff-rescaled}).}
as Ref.~\cite{Cebulla:2007ei}.

\begin{figure}[t!]
\centering
\begin{tabular}{ccc}
\includegraphics[height=5.1cm]{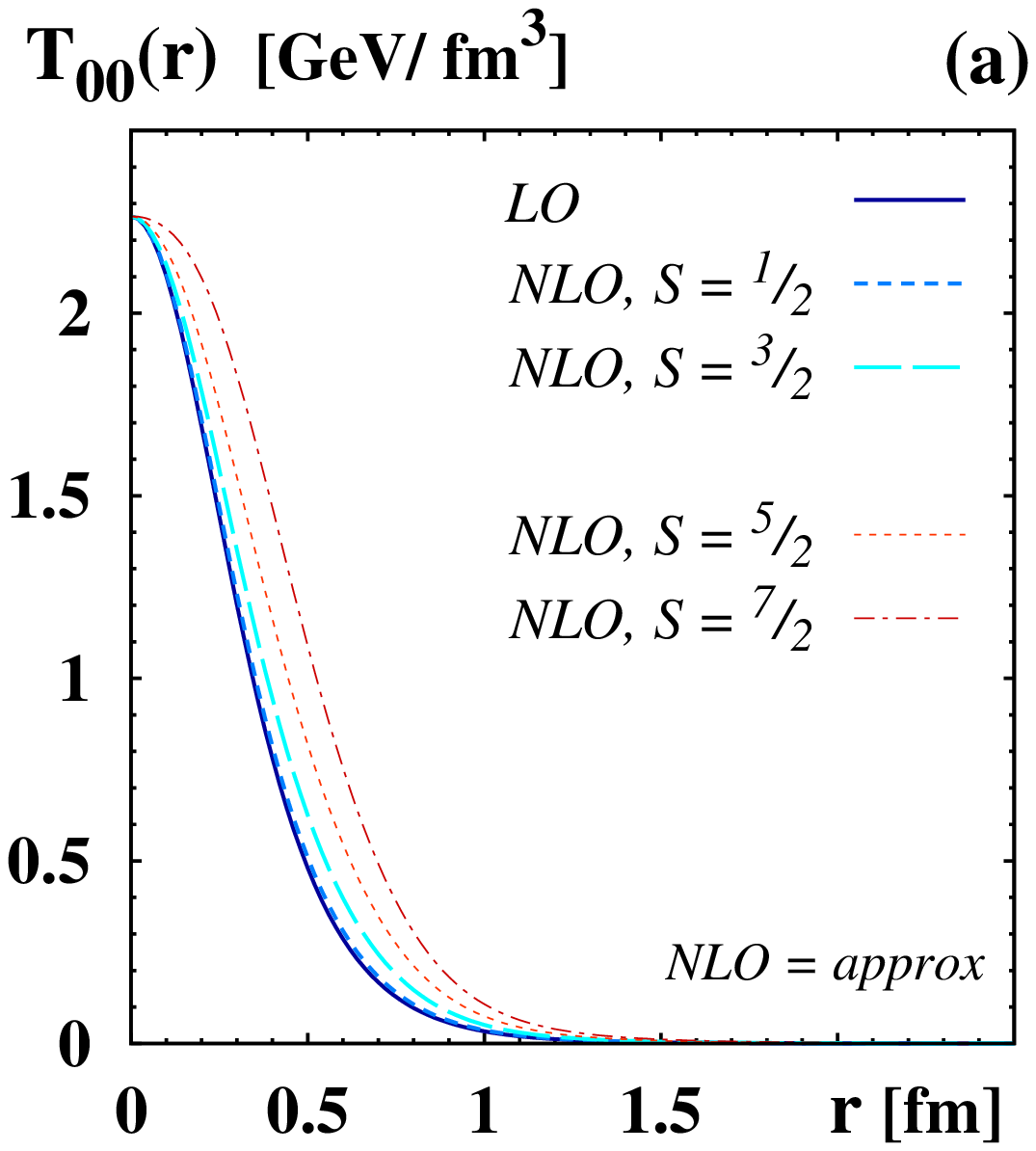} &
\includegraphics[height=5.1cm]{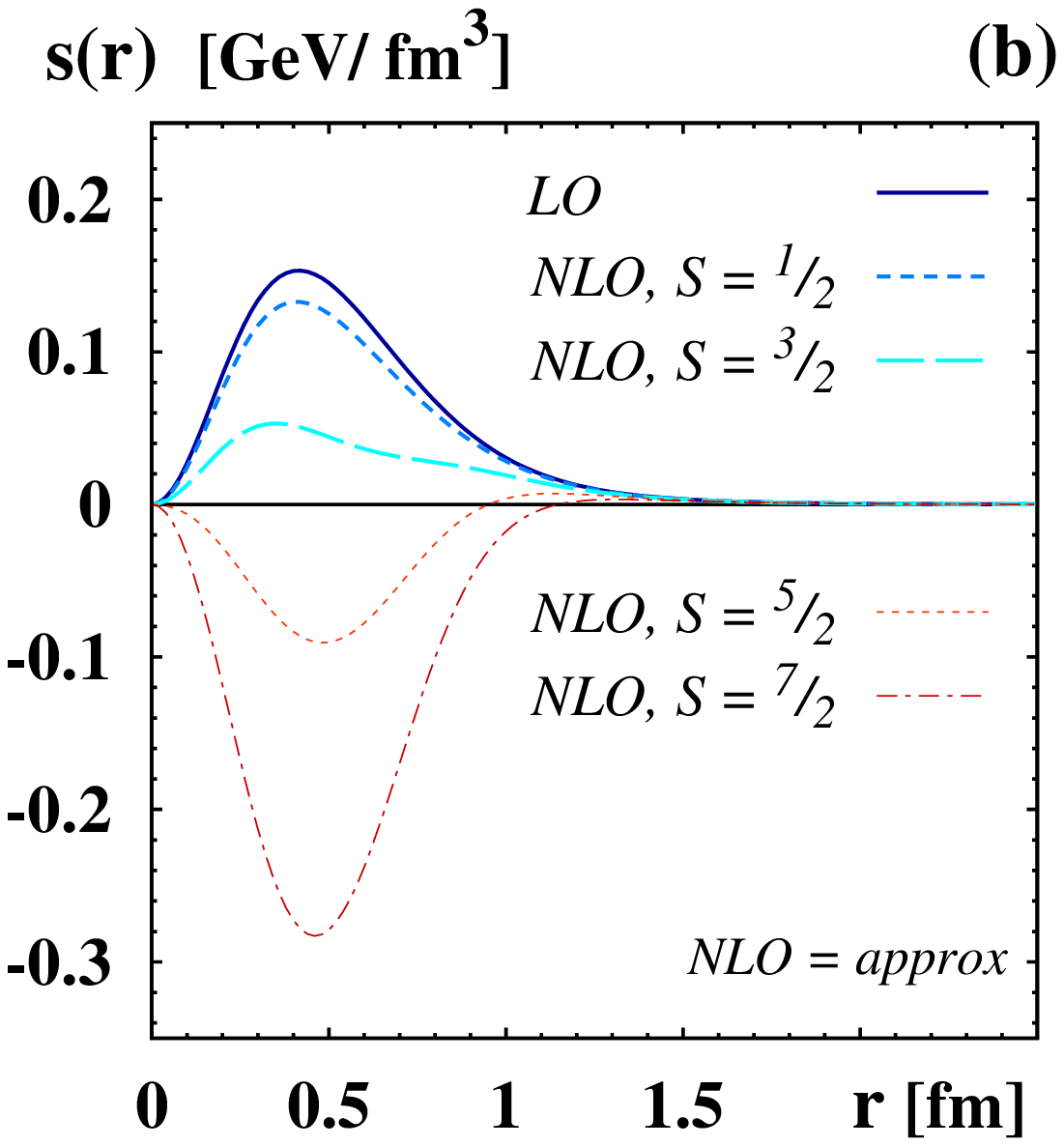} & 
\includegraphics[height=5.1cm]{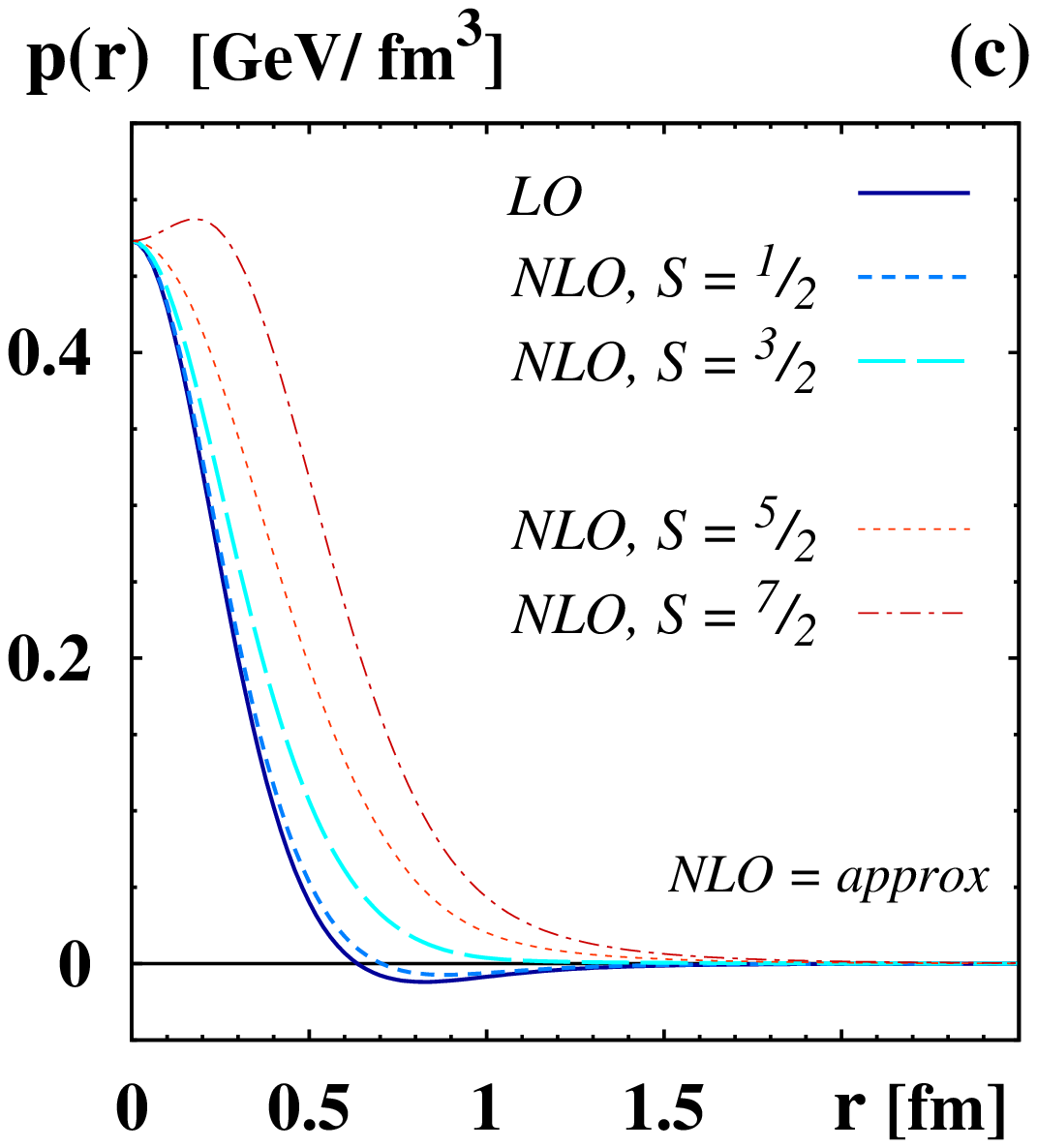} \\
\includegraphics[height=5.1cm]{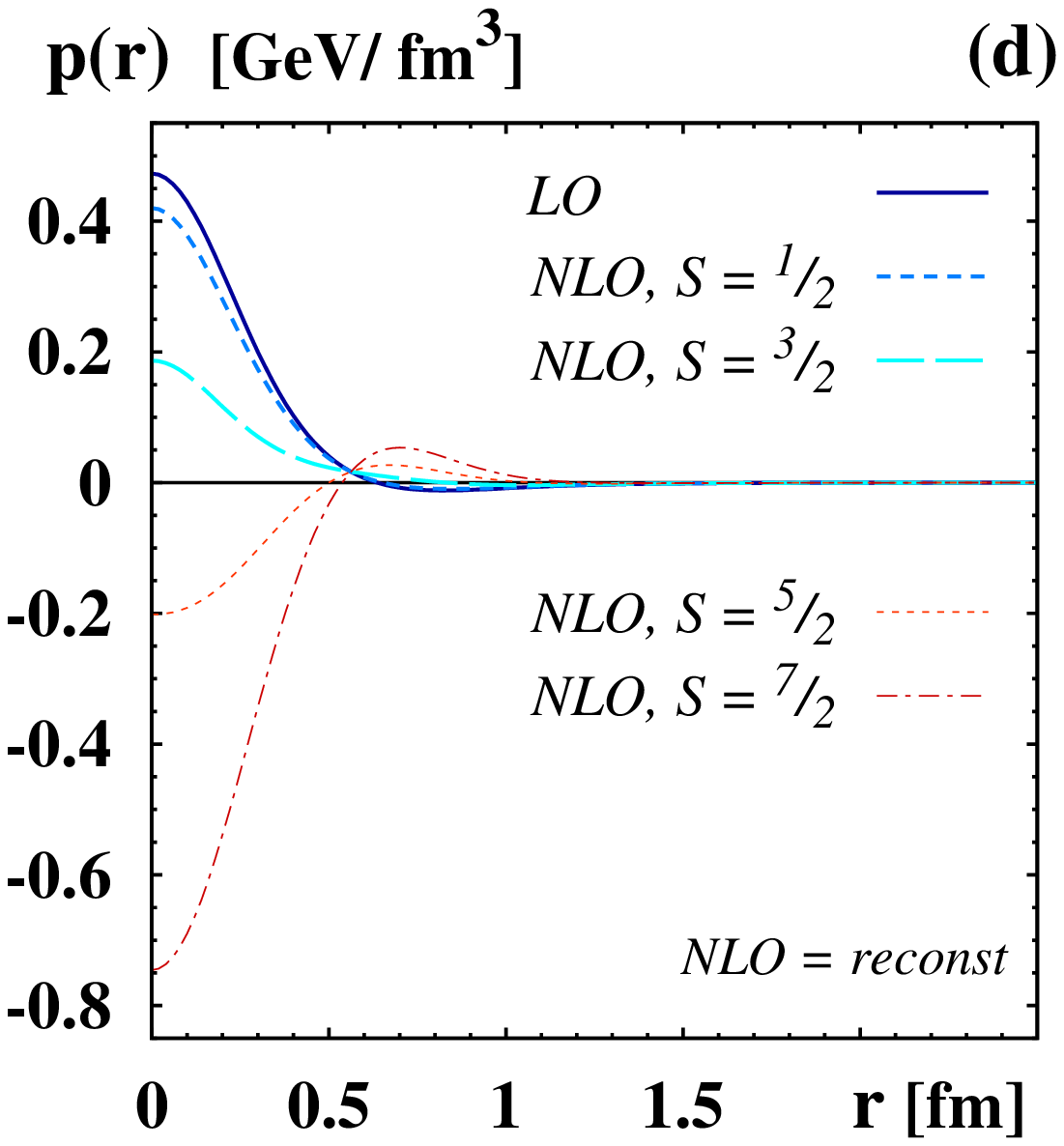} &
\includegraphics[height=5.1cm]{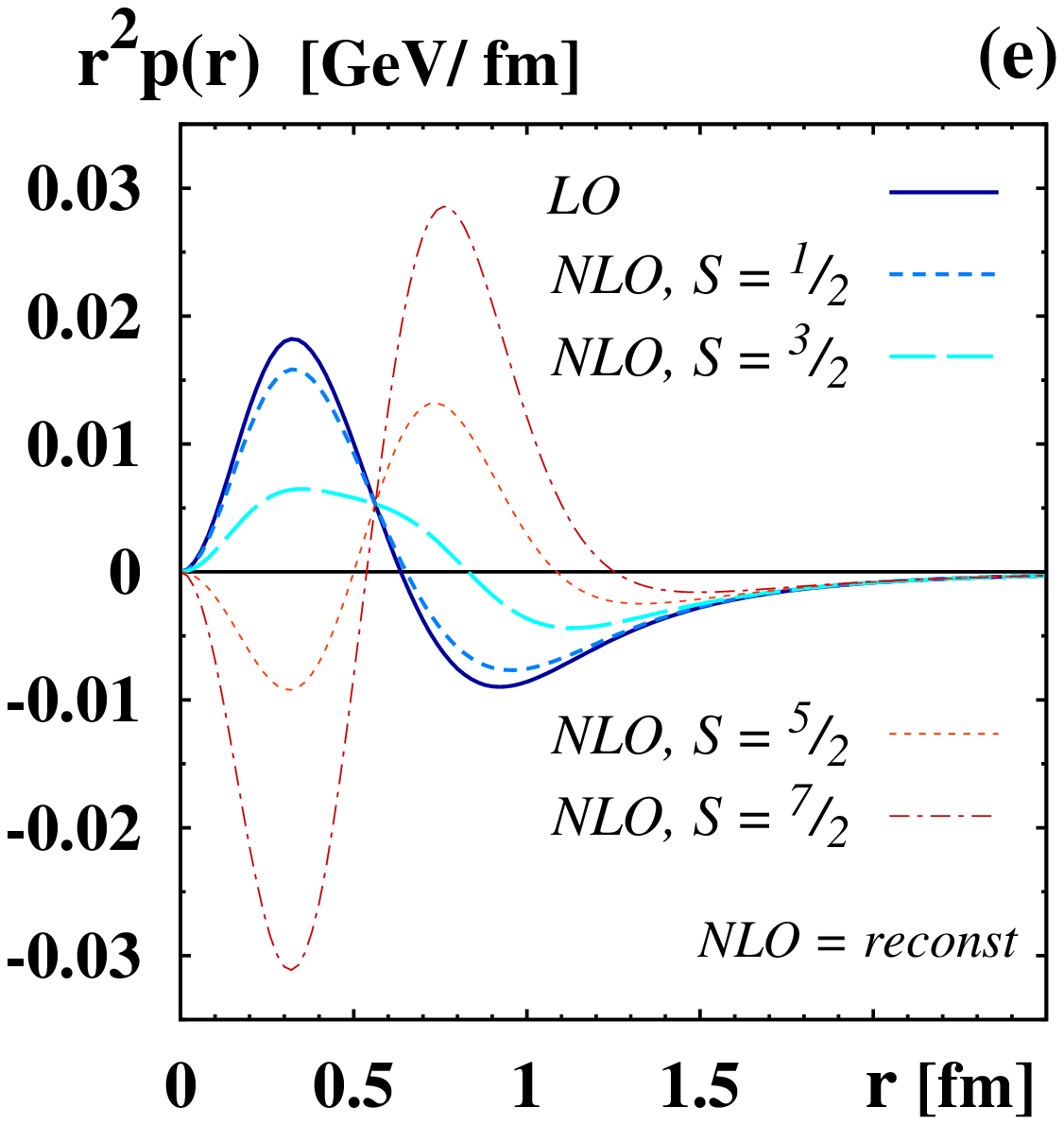} &
\includegraphics[height=5.1cm]{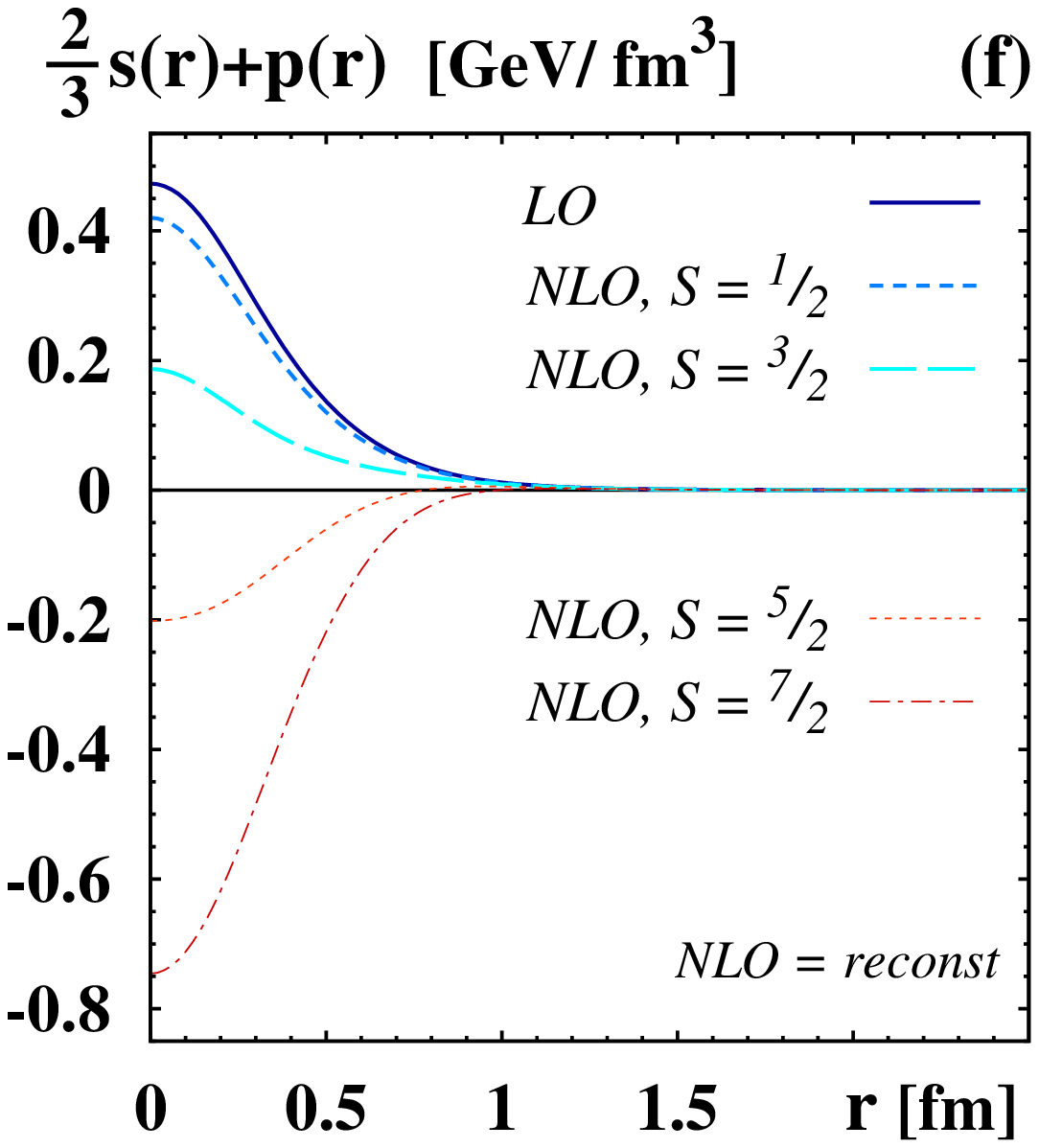} 
\end{tabular}
\caption{\label{FIG-01:densities-Skyrme}
	EMT densities from the Skyrme model as functions of $r$. The LO results 
	are valid for any $S=I$ in the large-$N_c$ limit. The estimates of NLO 
	corrections in the $1/N_c$-expansion are shown for states with the 
	quantum numbers $S=I=\frac12,\; \frac32,\; \frac52,\; \frac72$.
	The Figures show:
	(a) energy density $T_{00}(r)$,
	(b) shear forces $s(r)$,
	(c) pressure $p(r)$ with approximate NLO corrections,
	(d) $p(r)$ with NLO corrections reconstructed 
	according to (\ref{Eq:p-reconstructed}),  
	(e) same as Fig.~\ref{FIG-01:densities-Skyrme}d but 
	for $r^2p(r)$,
	(f) the local stability criterion (\ref{Eq:local-criterion-1}).
	The NLO results in the upper panel, 
	Figs.~\ref{FIG-01:densities-Skyrme}a--c,
	are estimated by simply evaluating the NLO expressions with the LO 
	soliton profile, which yields unacceptable results especially for 
	$p(r)$ which are at variance with Eq.~(\ref{Eq:von-Laue}). 
	The results in the lower panel,
	Figs.~\ref{FIG-01:densities-Skyrme}d--f, are obtained with the 
	pressure reconstructed according to (\ref{Eq:p-reconstructed}) 
	which satisfies Eq.~(\ref{Eq:von-Laue}). 
	Finally, Fig.~\ref{FIG-01:densities-Skyrme}f shows that the 
	results for $S=I=\frac12,\;\frac32$, which correspond to nucleon 
	and $\Delta$, comply with the local stability criterion 
	(\ref{Eq:local-criterion-1}). In contrast to this, states 
	with the exotic quantum numbers $S=I\ge 5/2$ do not satisfy 
	(\ref{Eq:local-criterion-1}), i.e.\ in this way the rotating soliton 
	approach explains why they are not realized in nature.}
\end{figure}

Let us begin the discussion with the energy density $T_{00}(r)$.
In Figs.~\ref{FIG-01:densities-Skyrme}a we show the LO result for 
$T_{00}(r)$ which is ``universal'' in the following sense. In LO of the
$1/N_c$ expansion nucleon and $\Delta$ are mass-degenerate, 
see Eq.~(\ref{Eq:Delta-N-mass-splitting}), i.e.\ both baryons have the
same energy density. More precisely, $T_{00}(r)$ is the same for the entire 
tower of light ground state baryons $S=I=\frac12,\;\frac32,\;\frac52,\;\dots$ 
including exotic quantum numbers. When NLO corrections due to soliton rotation 
are included, the mass depends on the $S=I$ quantum numbers according to 
Eq.~(\ref{Eq:Mrot}). In Fig.~\ref{FIG-01:densities-Skyrme}a we show 
the associated NLO results for $T_{00}(r)$ for the states 
$S=\frac12,\,\frac32,\,\frac52,\,\frac72$ which are normalized such that
$\int\di^3r\;T_{00}(r)$ yields the result in Eq.~(\ref{Eq:Mrot}) for the 
mass of the respective state. From Eq.~(\ref{Eq:Mrot}) it is clear that
higher spin states are heavier, and Fig.~\ref{FIG-01:densities-Skyrme}a 
shows that this is not due to higher density, but because the ``size'' 
of the system 	increases with $S$. 
This makes perfectly sense in a ``rigid-rotator'' approach, as the
NLO correction to $T_{00}(r)$ is proportional to the spin density
$\rho_J(r)$ which has the behavior $\rho_J(r)\propto r^2$ in soliton 
models \cite{Goeke:2007fp,Cebulla:2007ei}.\footnote{\label{Footnote:J}
	The spin density $\rho_J(r)$ is associated with the form 
	factor $J(t)$ in Eq.~(\ref{Eq:ff-of-EMT}) and related to 
	$T^{0k}(\bm{r},\bm{s})$ components of the static EMT 
	\cite{Polyakov:2002yz}.}
Remarkably, for $S=\frac12$ the LO and NLO results can hardly be 
distinguished on the scale of Fig.~\ref{FIG-01:densities-Skyrme}a.
This implies that for the nucleon the NLO corrections are moderate.
For higher spins $S=\frac32,\;\frac52,\;\frac72$ the NLO corrections become 
quickly more sizable. Nevertheless, $T_{00}(r)$ does not reveal anything 
unusual and looks equally plausible for all spin states. Other EMT densities,
namely $s(r)$ and $p(r)$, will turn out more insightful and give us a hint 
why the quantum numbers $S=I\ge\frac52$ are not observed in nature.

Next we investigate the distribution of shear forces. We recall that for a 
large nucleus, a situation which is well-described in the liquid drop model, 
the shear forces are given by $s(r)=\gamma\,\delta(r-R_N)$ where $R_N$ denotes
the radius of the nucleus and $\gamma$ the surface tension which can 
be inferred from the Bethe-Weizs\"acker formula \cite{Polyakov:2002yz}.
A realistic nucleus has no sharp edge and ``finite skin'' effects smear 
out the delta-function, but the liquid drop concept and consequences
from it remain valid \cite{Polyakov:2002yz,Guzey:2005ba}. However,
a single nucleon is much more diffuse, as can be seen from the LO 
result in Fig.~\ref{FIG-01:densities-Skyrme}b which shows a ``very 
strongly smeared out delta-function'' and an unambiguous definition
of the nucleon radius is not possible (although one may define 
certain mean square radii, see below Sec.~\ref{Subsec:some-results}). 
The NLO corrections are moderate in the case of 
the nucleon, see Fig.~\ref{FIG-01:densities-Skyrme}b.
For the $\Delta$ the NLO correction is much more sizable, where we
observe that $s(r)$ is clearly depleted and more strongly spread out. 
Thus, in the rotating soliton picture the $\Delta$ is a larger and
even more diffuse hadron than the nucleon. This is an intuitive 
and reasonable result. However, for the quantum numbers $S=I\ge \frac52$
we find a very different pattern: here the shear forces develop a node
being negative in the inner region and positive in the outer region.
A negative distribution of shear forces cannot be associated with 
a surface tension of a (however diffuse) particle, and in fact was
not observed in any of the theoretical studies performed so far
\cite{Goeke:2007fp,Goeke:2007fq,Cebulla:2007ei,Jung:2013bya,Kim:2012ts,
Jung:2014jja,Mai:2012yc,Mai:2012cx,Cantara:2015sna}.
The meaning of this result will become clear shortly.

Next we discuss the pressure. Let us first recall that only the LO
result in Fig.~\ref{FIG-01:densities-Skyrme}c satisfies the von
Laue condition in Eq.~(\ref{Eq:von-Laue}). This is so because the
condition (\ref{Eq:von-Laue}) is equivalent to the variational 
problem of minimizing the soliton mass $\delta M_{\rm sol} = 0$
\cite{Cebulla:2007ei}. If we added NLO corrections to $p(r)$ and 
evaluated them with a soliton profile obtained from the variational 
problem $\delta M_{\rm rot} = 0$ we of course would obtain results 
satisfying the von Laue condition (\ref{Eq:von-Laue}), but at the prize 
of unacceptable violations of chiral symmetry, see the discussion in 
Sec.~\ref{Sec-5a:Skyrme-with-Nc-corrections}.
If instead we use the LO soliton profile obtained from 
$\delta M_{\rm sol} = 0$ which preserves chiral symmetry, and evaluate 
$p(r)$ with NLO corrections ``added as a small perturbations'' as it
is customarily done, we obtain the ``approximate NLO'' results 
shown in Fig.~\ref{FIG-01:densities-Skyrme}c. These results
do not satisfy the von Laue condition.
Interestingly, on the scale of Fig.~\ref{FIG-01:densities-Skyrme}c 
the NLO correction to the nucleon looks moderate, and now we are
in the position to quantify this statement. The approximate NLO
result for the pressure does not satisfy the von Laue condition 
(\ref{Eq:von-Laue}) exactly, but does so ``approximately'' since
\be\label{Eq:von-Laue-nucleon-rot}
	\left[
	\frac{\int_0^\infty\di r\;r^2\,p(r)}
	     {\int_0^\infty\di r\;r^2\,|p(r)|}\right]_{\rm NLO, \, approx,\, S=\frac12}
 	= 0.30 \sim {\cal O}(N_c^{-1}).
\ee
In this sense the $1/N_c$ corrections in the nucleon case are moderate
and the von Laue condition remains ``satisfied'' within the accuracy 
one would expect after adding NLO corrections as a ``small perturbation.''
Another highly sensitive test is provided by evaluating the $D$-term 
$d_1$ from $s(r)$ and $p(r)$ according to 
Eqs.~(\ref{Eq:def-d1-pressure},~\ref{Eq:def-d1-shear}).
In LO we obtain the consistent result $d_{1,\rm LO}^p=d_{1,\rm LO}^s=-4.48$ where 
the subscripts indicate whether the value is obtained from $s(r)$ or $p(r)$. 
If one naively includes NLO corrections this equivalence is spoiled, and
we find $d_{1,\rm NLO}^p=-2.61$ vs $d_{1,\rm NLO}^s=-4.26$. The two results
agree within about $24\,\%\sim {\cal O}(N_c^{-1})$, 
i.e.\ also within the expected accuracy. 
It is important to stress that the approximate NLO result for $s(r)$ yields 
a $D$-term much closer to the LO result than the approximate NLO result for
$p(r)$. For $\Delta$ and higher spins 
states the NLO corrections 
to the pressure introduce a major qualitative change: the zero of $p(r)$
disappears,\footnote{
	Notice that this is for the optimized parameters of 
	\cite{Cebulla:2007ei}, see Footnote~\ref{Footnote:parameters}. 
	E.g.\ for the parameters of \cite{Adkins:1983ya} the NLO-effects 
	would be more drastic, and e.g.\ the node of $p(r)$ would disappear 
	already for $S=I=\frac12$  as shown in \cite{Cebulla:2007ei}.}
and we find a ``$100\,\%$ violation'' of the von Laue condition 
as measured analog to Eq.~(\ref{Eq:von-Laue-nucleon-rot}).
However, already the ``$30\,\%$ violation'' of the von Laue condition 
for the nucleon in Eq.~(\ref{Eq:von-Laue-nucleon-rot}) is not acceptable,
as this implies non-conservation of the EMT as explained in 
Sec.~\ref{Sec-5a:Skyrme-with-Nc-corrections}.

In order to obtain an acceptable estimate for the NLO corrections to the
pressure and construct a conserved EMT we have to reconstruct $p(r)$ from 
$s(r)$ according to Eq.~(\ref{Eq:p-reconstructed}). The ``reconstructed NLO'' 
results are shown in Fig.~\ref{FIG-01:densities-Skyrme}d. These results
satisfy the von Laue condition (\ref{Eq:von-Laue}) which we visualize
in Fig.~\ref{FIG-01:densities-Skyrme}e which shows the ``reconstructed NLO'' 
results for $r^2p(r)$. We again observe that the effects of NLO corrections
for the nucleon are small, and they are more sizable for $\Delta$. However,
both states $S=\frac12,\;\frac32$ exhibit the pattern of a stable physical
situation: positive $p(r)$ in the inner region, negative $p(r)$ in the
outer region, and exact balance according to the von Laue condition. The 
situation is fundamentally different for higher spin states $S\ge\frac52$:
here the reconstructed pressure also satisfies the von Laue condition,
but the signs are reversed and there is no balance of forces:
the negative $p(r)$ in the center corresponds to attractive forces which
are unbalanced, i.e.\ the inner part of the soliton collapses. At the same
time, the positive $p(r)$ in the outer region is also unbalanced, and 
the repulsive forces expel the outer part of the (too fast)
rotating soliton.

Let us also comment on the local stability criteria. All states
satisfy $T_{00}(r)\ge 0$ in agreement with (\ref{Eq:local-criterion-0}).
However, only the states with $S\le\frac32$ comply with the local
stability criterion (\ref{Eq:local-criterion-1}) while the states with 
$S\ge\frac52$ violate it as shown in Fig.~\ref{FIG-01:densities-Skyrme}f.
It is important to keep in mind, that this is a naive ``mechanical picture''
which is nevertheless very insightful. 
The rotating soliton approach predicts all states 
$S=I=\frac12,\;\frac32,\;\frac52,\;\dots$ on equal footing. It is
therefore remarkable that the approach itself explains that rotating 
solitons with $S=I\ge \frac52$ are artifacts of the rigid rotator 
quantization as they violate basic mechanical stability criteria
and therefore cannot correspond to physical states.

\subsection{Selected results}
\label{Subsec:some-results}

Having established a consistent scheme to estimate $1/N_c$ corrections 
to nucleon- and $\Delta$-properties in the Skyrme model, we end this 
section by stating some results of interest in the context of EMT 
densities. For the parameters used in this work, see
Footnote~\ref{Footnote:parameters}, we have in LO for the baryon
masses $M_\Delta=M_N=1085\;{\rm MeV}$. Including NLO corrections we 
obtain $M_N = 1159\;{\rm MeV}$ and $M_\Delta = 1452\;{\rm MeV}$.
Thus, the physical values of the masses are described within (20--30)\,$\,\%$
accuracy which is typical for this model \cite{Zahed:1986qz}.
Notice that soliton models generally tend to overestimate baryon masses 
to spurious contributions from rotational and translational zero-modes 
\cite{Pobylitsa:1992bk}.

In Table~\ref{Table:some-results} we summarize the Skyrme model predictions 
for selected EMT properties. Besides the $D$-term $d_1$ we include results
for the mean square radius of the energy density $\la r^2_E\ra$
and the mean square radius of the shear forces defined as
$\la r^2_s\ra = \int_0^\infty \di r\,r^2 s(r)/\int_0^\infty \di r\,s(r)$.
In addition, we also quote the results for the position $R_0$ at which
the pressure exhibits the node, i.e.\ $p(R_0) = 0$. 
The LO results are equal for nucleon and $\Delta$, but NLO corrections
remove this degeneracy. The NLO results for $d_1$ and 
$R_0$ are obtained with the reconstructed NLO-result for the pressure.

The results in Table~\ref{Table:some-results} show that NLO corrections
are small for the nucleon, and somewhat more sizable for $\Delta$.
In both cases they do not exceed 30$\,\%$ which one 
would naturally expect for $1/N_c$ corrections. With NLO corrections included,
the $D$-term of the nucleon is $-4.48$ and that of the $\Delta$ is $-3.31$.
Moreover, the $\Delta$ is larger than the nucleon which is quantified by the 
various radii in Table~\ref{Table:some-results}. This is an intuitive result
and in line with calculations of the electric mean square radius of $\Delta^+$ 
in models \cite{Ledwig:2008es} and lattice QCD \cite{Alexandrou:2008bn}.
The result for the $D$-term of the $\Delta$ in Table~\ref{Table:some-results} 
is to the best of our knowledge the first calculation of the $D$-term of the 
$\Delta$-resonance. Remarkably, also the $D$-term of the $\Delta$ is negative 
--- in agreement with theoretical calculations in other systems
\cite{Goeke:2007fp,Goeke:2007fq,Cebulla:2007ei,Jung:2013bya,Kim:2012ts,
Jung:2014jja,Mai:2012yc,Mai:2012cx,Cantara:2015sna}, see also
\cite{Pagels,Donoghue:1991qv,Ji:1997gm,Petrov:1998kf,Schweitzer:2002nm,Hagler:2003jd,Gabdrakhmanov:2012aa,Son:2014sna,Pasquini:2014vua}.

\begin{table}[h!]\centering
\begin{tabular}{l|c|c|c}
			&		& \multicolumn{2}{|c}{   } 	\\
			& LO 		& \multicolumn{2}{|c}{NLO} 	\\
			&    		& nucleon 	& $\Delta$-resonance \\ 
\ \hspace{1cm} \ & \ \hspace{2cm} \ & \ \hspace{2cm} \ & \ \hspace{2cm} \\ 
\hline
$\;\,d_1$		& -4.48 	& -4.25 	& -3.31   	\\
$\la r^2_E\ra^{1/2}$ 	& 0.74$\,$fm 	& 0.75$\,$fm	& 0.80$\,$fm	\\ 
$\la r^2_s\ra^{1/2}$ 	& 0.63$\,$fm 	& 0.64$\,$fm 	& 0.72$\,$fm	\\ 
$\;R_0$ 		& 0.64$\,$fm 	& 0.65$\,$fm 	& 0.83$\,$fm
\end{tabular}

\caption{\label{Table:some-results} 
	Selected EMT properties of the nucleon and $\Delta$ 
	from the Skyrme model. The results for the $D$-term $d_1$, 
	mean square radii of the energy density and shear forces, 
	$\la r^2_E\ra$ and $\la r^2_s\ra$, and the position $R_0$
	where the pressure distribution exhibits a node, refer 
	to LO (where the properties of 2 baryons are degenerate) 
	and to NLO of the $1/N_c$ expansion.}

\end{table}

\section{Charmonium-baryon bound states}
\label{Sec-6:bound-states}

After the general introduction to the EMT and its practical description
in the Skyrme model we are now in the position to discuss the effective
baryon-quarkonium potential.  

\subsection{\boldmath $V_{\rm eff}$ from Skyrme model}

\begin{figure}[t!]
\centering
\begin{tabular}{cc}
\includegraphics[height=5.1cm]{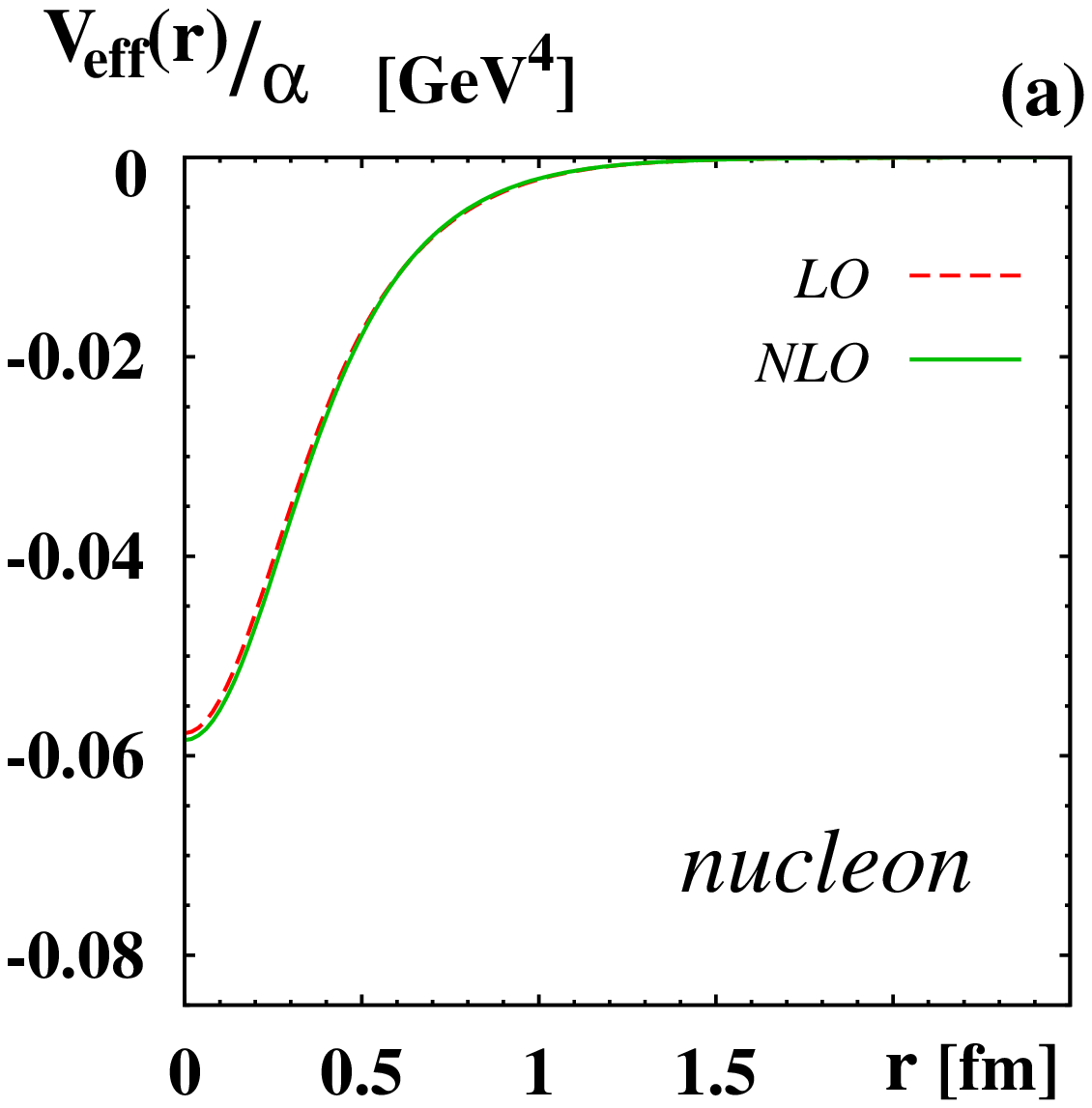} &
\includegraphics[height=5.1cm]{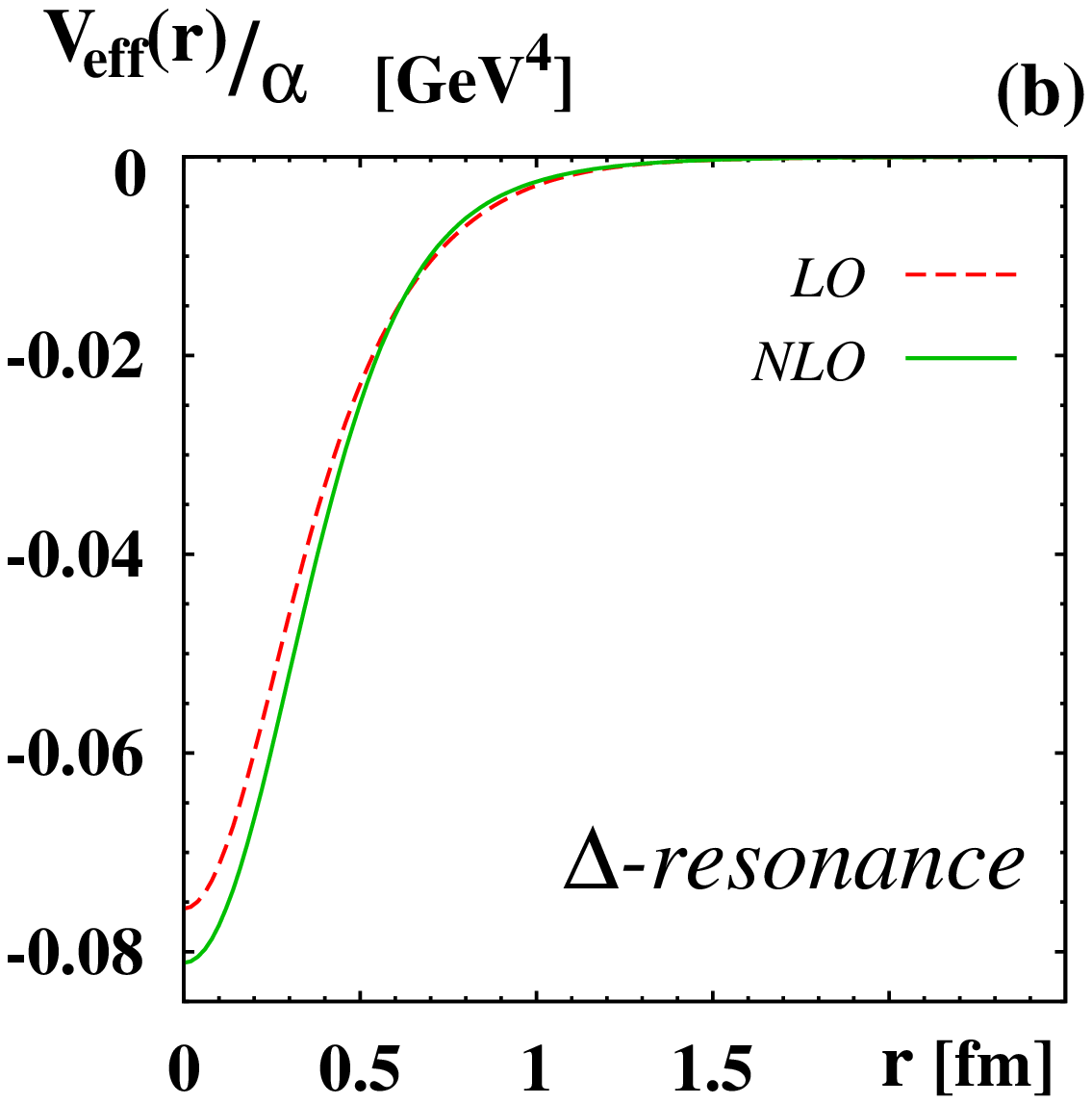}
\end{tabular}

\caption{\label{FIG-02:Veff-Skyrme}
	The effective potential $V_{\rm eff}(r)$ normalized with respect to the
	polarizability $\alpha$ for (a) nucleon, and (b) $\Delta$ 
	as function of $r$ in LO and NLO order of the large-$N_c$ expansion
	after the rescaling in Eq.~(\ref{Eq:Veff-rescaled}), i.e.\  
	$V_{\rm eff}(r)$ is normalized according to (\ref{Eq:Veff-norm}) 
	with respect to the physical value of the respective baryon mass.}
\end{figure}

Soliton models tend to overestimate baryon masses, and the Skyrme model 
(with our parameter fixing) is no exception in this respect, see previous 
section. In order to ensure a phenomenologically consistent description,
we rescale $V_{\rm eff}$ as follows
\be \label{Eq:Veff-rescaled}
	V_{\rm eff}(r) = -\alpha\frac{4\pi^2}{b }\left(\frac{g^2}{g_s^2}\right)
	\;\frac{M_{\rm physical}}{M_{\rm rot}}\;
	\left[ \nu\,T_{00}(0) - 3 p(r)\right],
\ee
so the effective potential is correctly normalized with respect 
to the physical value of the baryon mass in Eq.~(\ref{Eq:Veff-norm}).
Notice that one could also refrain from this step, and obtain the 
same results by redefining the value of $\alpha$.
The effective potentials for the nucleon and $\Delta$ obtained 
in this way are shown in Fig.~\ref{FIG-02:Veff-Skyrme}. 
We see that the effects of $1/N_c$ corrections are modest for the
nucleon, and somewhat more sizable for the $\Delta$-resonance.
The Skyrme model predictions for $V_{\rm eff}$ shown in 
Fig.~\ref{FIG-02:Veff-Skyrme} will be used in the following to 
investigate the dynamics in the nucleon- and $\Delta$-charmonium systems.

\subsection{Description of quarkonium-baryon bound states}

If a baryon-quarkonium bound state exists, its binding energy
$E_{\rm bind}<0$ follows from solving the non-relativistic Schr\"odinger 
equation
\be\label{Eq:Schroedinger}
	\left(-\frac{\bm{\nabla}^2}{2\mu}+V_{\rm eff}(r)-E_{{\rm bind}}
	\right)\Psi(\bm{r})=0,
\ee
where $\mu$ is the reduced mass\footnote{
	In the recent lattice QCD simulation \cite{Alberti:2016dru}
	it was investigated how the potential between a (infinitely heavy) 
	$q\bar{q}$-pair is modified if the heavy $q\bar{q}$-pair is placed 
	inside a light hadron. It was observed that the static potential,
	and consequently also quarkonium masses, are reduced by a few MeV.
	This ``medium effect'' is analog to the modification of e.g.\ 
	$\rho$-meson properties in nuclear environment, and should not 
	be confused with the binding energy of a heavy $q\bar{q}$-pair with 
	a light hadron, which is described by the effective interaction
	(\ref{Eq:Leff-1}). The results of \cite{Alberti:2016dru} imply
	that in our calculations we should use reduced charmonium masses
	instead of the physical ones. As other theoretical uncertainties 
	in our approach are more pronounced (heavy quark mass corrections, 
	$1/N_c$ corrections) we will neglect this small effect.}
in the channel of interest defined as 
$\mu^{-1} = M^{-1}_{\rm charmonium}+M^{-1}_{\rm baryon}$.

The Schr\"odinger equation \eqref{Eq:Schroedinger} can be 
conveniently rewritten using separation of 
variables and defining the radial function as follows
$\Psi(\bm{r}) = \Phi_{lm}(\vartheta,\varphi) u_l(r)/r$ with
boundary conditions $u_l(r) \propto r^{l+1}$ at small $r$,
and $u_{nl}(r)\to 0$ at large $r$ such that it can be
normalized as $\int_0^\infty\di r\,u_l^2(r)=1$. 
In principle there could be several bound states which should be
labeled accordingly by a radial quantum number, but we refrain 
from this to simplify notation.

Before starting the calculations let us recall that the shape
of $V_{\rm eff}(r)$ and its range, which can be defined e.g.\ 
in terms of $\la r_{\rm eff}^2\ra$ in \eqref{Eq:Veff-r2}, 
are determined by the model for the EMT densities and the
estimate for the parameter $\nu$ in \eqref{Eq:nu}. But the
overall normalization of the effective potential is basically
unconstrained due to the poor knowledge of the chromoelectric 
polarizabilities $\alpha$ for which only the rough guidelines
in Eq.~(\ref{Eq:alpha-guideline}) are available.
In practice it is therefore useful to treat $\alpha$ as a
free parameter, and vary it in a relatively wide region in order 
to determine whether bound states exist \cite{Eides:2015dtr}. 
Notice that the chromoelectric polarizability of $J/\psi$ 
in Eq.~(\ref{Eq:alpha-1S}) is so small and $V_{\rm eff}$ so shallow
that in our formalism no bound states of the nucleon and $J/\psi$ exist
--- even if we allow the numerical value of $\alpha(1S)$ to vary within
a reasonable range. In the following we will therefore focus on $\Psi(2S)$.
Hereby the lower bound derived in 
Sec.~\ref{Sec-3:condition-for-bound-state} will play a very helpful role.

\subsection{\boldmath 
Confirmation of $P_c(4450)$ as nucleon-$\psi(2S)$ bound state}

The states  $P_c(4380)$ and $P_c(4450)$ are observed to decay
in nucleon and $J/\Psi$, i.e.\ they have isospin $\frac12$ 
such that it is natural to consider the nucleon channel.
However, $J/\Psi$ itself {\it cannot} form bound states with
the nucleon, also because $M_N+M_{J/\Psi}=4035$~MeV is {\it smaller} 
than the mass of the lighter pentaquark $P_c(4380)$. Let us therefore 
focus here on nucleon-$\psi(2S)$ bound states.
In the following we will quote the numerical results obtained
from the Skyrme model in LO and NLO of the $1/N_c$ expansion,
and confront them with the results from the chiral quark-soliton
model ($\chi$QSM) reported in Ref.~\cite{Eides:2015dtr} which
also refer to LO of the large-$N_c$ limit.

In the eigenvalue problem \eq{Eq:Schroedinger} threshold bound 
states (i.e.\ states with infinitesimally small binding energies) 
emerge only if the chromoelectric polarizability is above a certain 
minimal value $\alpha_{\rm min}$ which depends on the orbital angular 
momentum quantum number $l$. We obtain for reduced mass of the
nucleon-$\Psi(2S)$ system (the numbers differ only slightly in the
nucleon-$J/\Psi$ system)
\ba\label{Eq:alpha-min-nucl-l-0}
	l = 0\,: && 
	\alpha > \alpha_{\rm min} = \begin{cases}
	5.1\;{\rm GeV}^{-3} & \mbox{Skyrme, LO,} \\
	5.0\;{\rm GeV}^{-3} & \mbox{Skyrme, NLO,} \\
	5.6\;{\rm GeV}^{-3} & \mbox{$\chi$QSM, Ref.~\cite{Eides:2015dtr},} 
	\end{cases} \\
\label{Eq:alpha-min-nucl-l-1}
	l = 1\,: && 
	\alpha > \alpha_{\rm min} = \begin{cases}
	23.8\;{\rm GeV}^{-3} & \mbox{Skyrme, LO,} \\
	23.5\;{\rm GeV}^{-3} & \mbox{Skyrme, NLO,} \\
	22.4\;{\rm GeV}^{-3} & \mbox{$\chi$QSM, Ref.~\cite{Eides:2015dtr}.} 
	\end{cases}
\ea
Bound states in the channels $l\ge 2$ would require polarizabilities
$\alpha > \alpha_{\rm min} = {\cal O}$(50--60)$\,;{\rm GeV}^{-3}$ and higher.
A comparison with the guideline (\ref{Eq:alpha-guideline}) for
$\alpha(1S)$ reveals that, even if it was energetically possible, 
$J/\Psi$ could not form bound states with the nucleon. 
However, for $\psi(2S)$, $\psi(3S)$, $\dots$ the minimal value of 
$\alpha$ in the $l=0$ channel is well below the perturbative guidelines 
in  Eq.~(\ref{Eq:alpha-guideline}) which means that the excited charmonia 
can form $s$-wave bound states with the nucleon. In the following we will 
focus on $\psi(2S)$ leaving the consideration of higher excited charmonia 
to future work. 
Notice that the result (\ref{Eq:alpha-min-nucl-l-0}) is in agreement with 
the bound derived in Sec.~\ref{Sec-3:condition-for-bound-state}.

Following the procedure of Ref.~\cite{Eides:2015dtr} we now 
determine which values of $\alpha(2S)$ would be required in order
to reproduce in the Skyrme model the exact binding energies 
$E_{\rm bind}=-176\,{\rm MeV}$ of $P_c^+(4450)$, and
$E_{\rm bind}=-246\,{\rm MeV}$ of $P_c^+(4380)$.
The binding energy of the heavier pentaquark state is exactly reproduced for 
\ba\label{Eq:alpha-P4450}
	P_c^+(4450): && 
	\alpha(2S) = \begin{cases}
	16.8\;{\rm GeV}^{-3} & \mbox{Skyrme, LO,} \\
	16.4\;{\rm GeV}^{-3} & \mbox{Skyrme, NLO,} \\
	17.2\;{\rm GeV}^{-3} & \mbox{$\chi$QSM, Ref.~\cite{Eides:2015dtr},} 
	\end{cases} 
\ea
while for $P_c^+(4380)$:
$\alpha(2S) = (19.6;\;19.1;\;20.2)\,{\rm GeV}^{-3}$ for
(Skyrme, LO; Skyrme, NLO; $\chi$QSM, Ref.~\cite{Eides:2015dtr}) respectively.
These values for $\alpha$ are in reasonable agreement with the perturbative 
estimate in Eq.~(\ref{Eq:alpha-guideline}), but in each case there is
only a single bound state. Therefore we have to choose which of the two
pentaquark states can be described in our formalism. The correct identification
can be made by considering the decay width.

The decay of a $\Psi(2S)$-nucleon bound state is driven by the potential
of the $2S\to 1S$ transition, which has the same ``universal shape'' as 
the $V_{\rm eff}$ responsible for the nucleon-$\psi(2S)$ binding mechanism,
but a significantly smaller normalization due to the small polarizability
relevant for the $2S \to 1S$ transition in Eq.~(\ref{Eq:alpha-guideline}). 
As this transition potential is relatively weak, one can use perturbation 
theory to estimate the decay width as follows \cite{Eides:2015dtr}
\be\label{Eq:Gamma}
	\Gamma = (4 \mu q)
	\left|\int_0^\infty dr r^2 u_l(r)V(r) j_l(q r)\right|^2.
\ee
Here $q$ is the center-of-mass momentum $q=\sqrt{2\mu E_R}$ 
where $E_R$ is the resonance energy and $\mu$ the reduced mass
of the decay products,
$j_l(z)$ is the spherical Bessel function, and $V_{\rm eff}(r)$ 
is the potential (\ref{Eq:Veff}) with the transitional polarizability 
$|\alpha(2S \to 1S)|=2\,{\rm GeV}^{-3}$ from phenomenological studies 
of $\psi^\prime\to J/\psi\, \pi\, \pi$ data \cite{Voloshin:2007dx}.
For the heavier pentaquark state we obtain in this way
\ba\label{Eq:width-P4450}
	P_c^+(4450): && 
	\Gamma = \begin{cases}
	17.0\;{\rm MeV} & \mbox{Skyrme, LO,} \\
	15.1\;{\rm MeV} & \mbox{Skyrme, NLO,} \\
	11.2\;{\rm MeV} & \mbox{$\chi$QSM, Ref.~\cite{Eides:2015dtr},} 
	\end{cases} 
\ea
which is in reasonable agreement with the observed width of $P_c^+(4450)$ 
quoted in Table~\ref{tab:1}.
For $P_c^+(4380)$ our formalism would yield a similarly narrow width 
$\Gamma = (21.3;\;18.8)\,{\rm GeV}^{-3}$ for (Skyrme, LO; Skyrme, NLO),
but the experimental result is an order of magnitude larger,
see Table~\ref{tab:1}.
Thus, the $s$-wave nucleon-$\psi(2S)$ bound state found in our 
approach is clearly identified with the heavier and narrower
state $P_c^+(4450)$. The lighter but broader resonance $P_c(4380)$ 
does not appear to be a nucleon-$\psi(2S)$ bound state.

Our results confirm the interpretation of $P_c^+(4450)$ as a 
nucleon-$\Psi(2S)$ state \cite{Eides:2015dtr}. A remarkably
consistent and robust picture emerges from our calculation 
and the comparison to the results of \cite{Eides:2015dtr}.
For a rather well well-constraint value of the chromoelectric 
polarizability of
\be\label{Eq:alpha-2S-our}
	\alpha(2S) = (16\mbox{--}17)\,{\rm GeV}^{-3}
\ee
two different models of the nucleon, the $\chi$QSM used in 
\cite{Eides:2015dtr} and the Skyrme model used in this work,
predict a naturally narrow bound state in the $l=0$ channel 
of the effective potential (\ref{Eq:Leff-1}) which can be 
identified with $P_c^+(4450)$. 
The results based on the $\chi$QSM in 
Eqs.~(\ref{Eq:alpha-min-nucl-l-0},~\ref{Eq:width-P4450})
refer to the LO of the $1/N_c$ expansion and are systematically 
closer to the LO results in the Skyrme model. Comparing the
numbers from LO and NLO within the Skyrme model shows that the 
predictions are also robust with respect to $1/N_c$ corrections.

The approach predicts the following quantum numbers for $P_c^+(4450)$.
The vector-meson $\psi(2S)$ has $J^P=1^-$ and nucleon has $J^P=\frac12{ }^+$. 
Considering that it is a bound state in the $l=0$ channel, 
the parity of $P_c^+(4450)$ is predicted to be negative.
The approach predicts actually not one but two states 
with spins $\frac12$ and $\frac32$ with a mass difference,
caused by hyperfine splitting due to quarkonium-nucleon spin-spin 
interaction, which is suppressed in the heavy quark mass limit
\cite{Eides:2015dtr}. 
The quantum numbers $J^P=\frac32{ }^-$ are consistent with experiment,
see Table~\ref{tab:1}. 

A comment regarding the spin-parity assignment for $P_c^+(4450)$ is in order. 
The result preferred by the LHCb analysis is 
$\frac52{ }^+$ \cite{Aaij:2015tga}. 
The assignments $\frac52{ }^-$ and $\frac32{ }^-$ are
within respectively 1-sigma and 2.3-sigma of the preferred fit,
i.e.\ also compatible with data, while assignments like $\frac12{ }^\pm$ 
or $\frac72{ }^\pm$ are disfavored at 5-sigma level \cite{Aaij:2015tga}.
Of course, one should keep in mind that the LHCb analysis did not test 
the hypothesis that the structure around $4450\,{\rm GeV}$ could consist of 
two nearly degenerate states with $J^P={\frac32}{ }^-$ and $J^P={\frac12}{ }^-$
as predicted in the current approach. It would be very interesting to perform
such a test.

\subsection{Prediction of a charmonium-$\Delta$ bound state}

In Ref.~\cite{Eides:2015dtr} it was argued that not only the
nucleon but also other baryons could potentially form bound states 
with charmonia via the effective interaction (\ref{Eq:Leff-1}).
With the results obtained on the EMT of $\Delta$ in
Sec.~\ref{Sec-5:Skyrme} we are in the position to investigate the 
question whether $\Delta$ can form bound states with charmonia. 
We will denote the possible charmonium-$\Delta$ bound states
as $\Pnew$.

In order for the effective charmonium-$\Delta$ potential $V_{\rm eff}$ 
to be strong enough to form bound states in the channels with angular
momentum $l$ the polarizabilities must be above the following minimal values
\ba\label{Eq:alpha-min-delta-l-0}
	l = 0\,: && 
	\alpha > \alpha_{\rm min} = \begin{cases}
	3.1\;{\rm GeV}^{-3} & \mbox{Skyrme, LO,} \\
	3.0\;{\rm GeV}^{-3} & \mbox{Skyrme, NLO,} 
	\end{cases} \\
\label{Eq:alpha-min-delta-l-1}
	l = 1\,: && 
	\alpha > \alpha_{\rm min} = \begin{cases}
	14.7\;{\rm GeV}^{-3} & \mbox{Skyrme, LO,} \\
	14.0\;{\rm GeV}^{-3} & \mbox{Skyrme, NLO,} 
	\end{cases} \\
\label{Eq:alpha-min-delta-l-2}
	l = 2\,: && 
	\alpha > \alpha_{\rm min} = \begin{cases}
	33.6\;{\rm GeV}^{-3} & \mbox{Skyrme, LO,} \\
	31.9\;{\rm GeV}^{-3} & \mbox{Skyrme, NLO.} 
	\end{cases}
\ea
Confronting these results with the guideline (\ref{Eq:alpha-guideline}) 
for $\alpha(1S)$ reveals that $J/\Psi$ cannot bind with $\Delta$.
However, $\psi(2S)$ could form bound states with $\Delta$ in the
$l=0$ and $l=1$ channels if we rely on our own estimate 
(\ref{Eq:alpha-2S-our}). This is supported by the model-independent 
bound in Sec.~\ref{Sec-3:condition-for-bound-state}).

In order to proceed with the calculation of possible bound states of
$\Delta$ and $\psi(2S)$ we will fix $\alpha(2S)$ at the values in
Eq.~(\ref{Eq:alpha-P4450}) which were required to explain $P_c(4450)$ 
as a nucleon-$\psi(2S)$ bound state, and use the respective LO and NLO
predictions from the Skyrme model for the effective potential as 
shown in Fig.~\ref{FIG-02:Veff-Skyrme}. In this way we obtain the
prediction that there is a single bound state in the $l=0$ channel 
with the binding energy and mass 
\be\label{Eq:prediction-Psi-Delta-mass-l-0}
	l = 0: \;\;\;
	E_{\rm bind} = \begin{cases}
	-370\;{\rm MeV} & \mbox{Skyrme, LO,} \\
	-430\;{\rm MeV} & \mbox{Skyrme, NLO,} 
	\end{cases}
	\; \; \; \Leftrightarrow \; \; \; 
	M = \begin{cases}
	4.54\;{\rm GeV} & \mbox{Skyrme, LO,} \\
	4.49\;{\rm GeV} & \mbox{Skyrme, NLO.} 
	\end{cases}
\ee
Estimating the width of the new state according to Eq.~(\ref{Eq:Gamma}) 
we obtain 
\be\label{Eq:prediction-Psi-Delta-width-l-0}
	l = 0: \;\;\;
	\Gamma = \begin{cases}
	55\;{\rm MeV} & \mbox{Skyrme, LO,} \\
	68\;{\rm MeV} & \mbox{Skyrme, NLO.} \\
	\end{cases} 
\ee
We again observe that the predictions are numerically very stable with 
respect to model details like effects of $1/N_c$ corrections.
The parity of this new state is negative and isospin is $\frac32$.
By applying the arguments of \cite{Eides:2015dtr} regarding the
spin assignment we predict that there are three states with 
$J=\frac12,\; \frac32,\; \frac52$ which are mass-degenerate 
modulo heavy quark mass corrections.
In the following we will refer to this new state as $\Pnew(4500)$.

Let us now turn to the $l=1$ channel. In this case our estimated result
for $\alpha(2S)$ in Eq.~(\ref{Eq:alpha-2S-our}) is much closer to the 
minimal value of $\alpha$ in Eq.~(\ref{Eq:alpha-min-delta-l-1}), but
it is clearly above it and a single bound state exists although it 
is loosely bound. The calculation yields 
\be\label{Eq:prediction-Psi-Delta-mass-l-1}
	l = 1: \;\;\;
	E_{\rm bind} = \begin{cases}
	-25\;{\rm MeV} & \mbox{Skyrme, LO,} \\
	-36\;{\rm MeV} & \mbox{Skyrme, NLO,} 
	\end{cases}
	\; \; \; \Leftrightarrow \; \; \; 
	M = \begin{cases}
	4.89\;{\rm GeV} & \mbox{Skyrme, LO,} \\
	4.88\;{\rm GeV} & \mbox{Skyrme, NLO,} 
	\end{cases}
\ee
The estimate of the width of the new state according to Eq.~(\ref{Eq:Gamma}) 
yields $\Gamma_2 = (5.0;\;9.5)\,{\rm GeV}^{-3}$ for (Skyrme, LO; Skyrme, NLO),
but this is not the total width. This state is below the threshold for
$\Psi(2S)$--$\Delta$ production, but it is above the threshold for 
$\Psi(2S)$-nucleon-pion production. This means that the $\Delta$ in
this bound state has the phase space to decay to the pion-nucleon
final state without ``waiting'' for the transition of $\Psi(2S)$ to
$J/\Psi$ to occur. The dominant decay mode for this resonance is
therefore $\Pnew(4900)\to \Psi(2S)\,N\,\pi$ with a partial decay
width $\Gamma_1\sim 150\,{\rm MeV}\gg\Gamma_2$ which is determined by the
width of the $\Delta$. For the total decay width we therefore predict
\be\label{Eq:prediction-Psi-Delta-width-l-1}
	l = 1: \;\;\;
	\Gamma = \Gamma_1 + \Gamma_2 
	\gtrsim 150\;{\rm MeV} \;\;\;\mbox{Skyrme, LO \& NLO}.
\ee
The parity of this $p$-wave state is positive, 
and isospin is $\frac32$. The possible spins are in the range
$\frac12 \le J \le \frac72$ following from combining spin $1$ of 
$\Psi(2S)$, spin $\frac32$ of $\Delta$, and orbital angular momentum $l=1$.
The different spin states again are mass-degenerate in the heavy quark mass 
approximation \cite{Eides:2015dtr}.
In practice, heavy quark mass corrections \cite{Eides:2015dtr} could 
shift the masses of (some of) these states into the 
$\Psi(2S)$--$\Delta$ continuum, i.e.\ they could be presumably even broader. 
Due to the proximity to the threshold $\Psi(2S)$--$\Delta$ threshold the 
theoretical uncertainties of this predictions could be larger than in
in the $l=0$ channel.

The general reason why a prospective $l=1$ state appears in the 
$\Psi(2S)$-$\Delta$ system, but not in the $\Psi(2S)$-nucleon system, 
is related to the larger mass of the $\Delta$ which enters the normalization 
of the potential in Eq.~(\ref{Eq:Veff-norm}). For heavier baryons 
lower values for $\alpha_{\rm min}$ are required to form bound states, see
Eqs.~(\ref{Eq:alpha-min-nucl-l-0},~\ref{Eq:alpha-min-nucl-l-1}) 
vs (\ref{Eq:alpha-min-delta-l-0},~\ref{Eq:alpha-min-delta-l-1}).
This is supported by the
model-independent bounds of Sec.~\ref{Sec-3:condition-for-bound-state}.
As a consequence heavier baryons in general form more easily bound states
with charmonia, perhaps even with bottonia.

\newpage
\section{Possible ways to observe $\Pnew$}
\label{Sec-7:how-to-observe}

In this section we will discuss possible ways to observe the 
newly predicted charmonium-$\Delta$ bound states $\Pnew$.

\subsection{$\Pnew$ and its $SU(3)$ partners in decays of bottom baryons}

The pentaquarks $P_c$ were observed in the decay 
$\Lambda_b^0 \to J/\Psi\,p\,K^-$ \cite{Aaij:2015tga} and their existence 
is supported by studies of the decay $\Lambda_b^0 \to J/\Psi\,p\,\pi^-$ 
\cite{Aaij:2016ymb}. These weak decays correspond to $b\to c \bar c s$ 
and $b\to c \bar c d$ transitions correspondingly. The second transition 
is Cabbibo suprsessed. 
In the following we will discuss both types of transitions.

\subsubsection{Transitions with $\Delta S=-1$  and $\Delta I=0$}

In this case the decay
$\Lambda_b^0 \to \Delta^+_c K^- \to J/\Psi\, p\, \pi^0\, K^- $ is forbidden 
and hence the $J/\Psi\, N\,\pi\, \bar K$ final state in the decay of 
$\Lambda_b^0$ is not suitable for the search of $\Pnew$.  However, if one 
considers the final state  $J/\Psi\, N\,\pi\, \bar K$ in decays of the 
isospin-1 baryons $\Sigma_b$ the isospin-3/2 pentaquarks $\Pnew$ can be 
found there. Presumably the most easily detectable modes 
(no neutral particles in the final state) are:

\ba
&&\Sigma_b^-\to \Pnew^0\, K^-\to J/\Psi\, p\, \pi^-\, K^-,\\
&&\Sigma_b^+\to \Pnew^{++}\, K^-\to J/\Psi\, p\, \pi^+\, K^-.
\ea
We note that $\Delta S=-1$ decays of $\Xi_b\to J/\Psi\, Y\, \bar K$ (where $Y$ is a baryon with $S=-1$ from the octet or decuplet)
are suitable for search of strange flavour $SU(3)$ partners\footnote{
	On general grounds we expect that $\Psi(2S)$ can be stronger 
	(than to nucleon and $\Delta$) bound to strange members of 
	the octet and the decuplet.} 
of $P_c$ and $\Pnew$ pentaquarks. Very interesting possibility to search 
for $P_{\Omega c}$ (the $S=-3$ decuplet partner of the pentaquark $\Pnew$) is 
provided by studies of the decay $\Xi_b^0 \to J/\Psi\, \Omega^-\, K^+$. 

\subsubsection{Transitions with $\Delta S=0$  and $\Delta I=\frac 12$}

For such Cabbibo suppressed transition $\Pnew$ can be searched in $\Lambda_b^0$
decays with the final state $J/\Psi\, N\,\pi\, M$ (where $M$ is a isospin-1 
meson, e.g. $\pi$-meson). In decays of $\Sigma_b$ with the same final state 
$\Pnew$ shows up also for the case where $M$ is a isospin-0 meson,
e.g. $\eta$-meson. In decays of $\Xi_b$ the pentaquarks $\Pnew$ can be 
searched in the final state $J/\Psi\, N\,\pi\, \bar K$. Note that 
the strange octet and decuplet partners of $P_c$ and $\Pnew$ can be 
looked for in the same decay mode.

\subsection{$\Pnew$ formation in photon and meson scattering on the nucleon}

In Refs.~\cite{Wang:2015jsa,Kubarovsky:2015aaa,Karliner:2015voa} it was 
suggested to search for $P_c$ pentaquarks through its formation
in the process $\gamma+p\to P_c\to J/\Psi + p$. One might think that the 
search for $\Pnew$ could be possible in the formation experiment
like $\gamma+p\to \Pnew\to J/\Psi+N+\pi$. However, here we expect that 
the $\gamma\, N\, \Pnew$-vertex is much smaller than the analogous
$\gamma\, N\, P_c$vertex because the former involves isospin $1/2\to 3/2$ 
transition and hence the overwhelming (see discussion in 
\cite{Kubarovsky:2015aaa}) vector dominance $\gamma\to J/\Psi$ transition 
does not contribute. It seems that the more favourable $\Pnew$ formation 
process is:
 \be
 \gamma+p\to \Pnew+\pi\to J/\Psi+N+\pi+\pi.
 \ee  
In such process the $\gamma\to J/\Psi$ transition makes large contribution.
The minimal photon energy in a fixed target experiment needed to
produce $\Pnew(4500)$ is 11 GeV, i.e.\ above the energies accessible 
in the Gluex Experiment at Jefferson Lab.

In Ref.~\cite{Kim:2016cxr} the formation of $P_c$ pentaquarks was considered 
in pion induced processes $\pi+N\to P_c\to J/\Psi+N$. It was shown
that the signal cross-section is of order 1~nb. Obviously, the $\Pnew$ 
pentaquark formation in $\pi+N\to \Pnew\to J/\Psi+N+\pi$ is of 
similar size, but, probably with smaller background.
These reactions could be studied in the charm spectroscopy
program at J-PARC, where pion beams with energies up to 20 GeV are
available \cite{Shirotori:2015eqa}.

\section{Conclusions}
\label{Sec-8:conclusions-outlook}

In this work we made use the formalism of Ref.~\cite{Eides:2015dtr} where 
the narrow $P^+_c(4450)$ state was interpreted as a nucleon-$\psi(2S)$ 
$s$-wave bound state with $J^P=\frac32{ }^-$. In the framework of this
formalism we derived a general lower 
bound which the charmonia chromoelectric polarizabilities must satisfy 
such that charmonium-baryon bound states can exist, and shown in 
model-independent way that $\psi(2S)$ can form $s$-wave bound 
states with nucleon and $\Delta$.

Using the Skyrme model for the densities of the EMT we have confirmed 
in detail the calculations from Ref.~\cite{Eides:2015dtr} which were 
based on a different model of the nucleon (chiral quark soliton model). 
The emerging picture for $P^+_c(4450)$ as a nucleon-$\psi(2S)$ bound state 
is very robust and insensitive to details of the underlying models. 
A particulary important aspect of model dependence is related to $1/N_c$ 
corrections. We have shown that the conclusions and numerical details of the 
calculations regarding $P^+_c(4450)$ are unaffected by $1/N_c$ corrections. 

As an interesting by-product of our study, we have shown how to construct 
a conserved EMT when a theory or model cannot be solved exactly and e.g.\
$1/N_c$ corrections must be included as a small perturbation. The soliton
approach describes baryons with spin and isospin quantum numbers 
$S=I=\frac12,\;\frac32,\;\frac52,\;\dots\,$ in the large-$N_c$ limit.
We have shown that, when $1/N_c$ corrections are included, it is
possible to construct a conserved EMT with densities which obey
fundamental stability criteria only for $S=I=\frac12,\;\frac32$
which correspond to nucleon and $\Delta$.
But for $S=I\ge\frac52$ the $1/N_c$ corrections are too destabilizing,
explaining why such states are not observed in nature.

We have investigated whether charmonia can bind with $\Delta$ to produce 
results which could allow us to further test this approach.
We have shown that the approach predicts a negative-parity $s$-wave bound 
state in the $\Delta$-$\psi(2S)$ channel with a mass around $4.5\,{\rm GeV}$ 
and width around $70\,{\rm MeV}$. It also predicts a broader 
positive-parity $p$-wave resonance around $4.9\,{\rm GeV}$ with width of 
the order of $150\,{\rm MeV}$. Each of these states contains several
spin states with mass-differences (caused by hyperfine splitting due to 
quarkonium-baryon spin-spin interaction) which are suppressed in the 
heavy quark mass limit.

An important question concerns how to observe these new pentaquark states.
We have examined suitable weak decays of bottom-baryons $\Lambda_b^0$, 
$\Sigma_b$, $\Xi_b$ where the new pentaquark states $\Pnew$ could be 
observed. We have also discussed how the $\Pnew$ could be observed in
photon-nucleon or pion-nucleon scattering reactions.

An important future direction is to extend the formalism to include 
charmonium-hyperon bound states. As hyperons are heavier the formation 
of such bound states is more favorable to the nucleon case.
Particularly interesting new pentaquark states would include 
charmonium-$\Omega$ bound states $P_{\Omega c}$, which include the
$S=-3$ decuplet partner of $\Pnew$, have the minimal content 
$sssc\bar{c}$, and could be detected in weak decays of 
$\Xi_b^0 \to J/\Psi\, \Omega^-\, K^+$. The properties of these and 
other hyperon-charmonium bound states will be addressed in future work.
As they scale with the size of the system, the chromoelectric 
polarizabilities of bottomia are too small and the resulting effective 
interactions too weak to form nucleon-bottomium bound states.
An interesting open question concerns the possibility whether the
heavier hyperons may form bound states with bottomia. This is another
interesting topic to be explored in future studies.

\

\acknowledgments
M.V.P.\ is grateful to M.~Eides, Yu.~Panteleeva and V.~Petrov for 
many illuminating discussion.
This work was supported in part by the National Science Foundation 
(Contract No.~1406298), and the Deutsche Forschungsgemeinschaft
(Grant VO 1049/1).

\appendix

\section{Chiral properties of EMT densities}
\label{App-A:large-r}

In this Appendix we review the large-$r$ behavior of the EMT
densities $T_{00}(r)$, $s(r)$, $p(r)$ derived from soliton models
in the large-$N_c$ limit.
The chiral soliton fields are described in terms of profiles $P(r)$
\cite{Goeke:2007fp,Cebulla:2007ei}. Although the dynamics of the 
different models is much different, chiral symmetry uniquely dictates 
that the profiles exhibit at asymptotic distances, in practice at 
$r\gtrsim (1\mbox{--}2)\,{\rm fm}$, the following behavior 
\be\label{Eq:profile-large-r}
	P(r) = \frac{2R_0^2}{r^2}\,(1 + m_\pi r)\,\exp(-m_\pi r) + \dots
\ee
where the dots indicate subleading terms. This behavior is universal,
i.e.\ valid for all (light) baryons in the large-$N_c$ limit.
The ``soliton size'' $R_0$ is a characteristic and in general 
model-dependent length-scale in the respective model. 
In the chiral limit, however, it can be related model-independently
to the axial-coupling constant of the nucleon and the pion decay
constant, see Eq.~(\ref{Eq:R0-gA}).

In the $\chi$QSM and the Skyrme model the large-$r$ 
behavior of the EMT densities can be computed analytically
\cite{Goeke:2007fp,Cebulla:2007ei}. Retaining only the leading chiral 
contributions, one obtains from (\ref{Eq:profile-large-r}) the results
\cite{Cebulla:2007ei}
\begin{subequations}\label{Eq:EM-FF-large-r}
\begin{align}
	\label{Eq:T00-large-r}
	T_{00}(r)& =\phantom{-}\, \frac12\;\frac{F_\pi^2\;R_0^4}{r^6}\;
		(6 + 12 m_\pi r+11 m_\pi^2 r^2+\;6 m_\pi^3 r^3 \,+2 m_\pi^4 r^4)
		\,e^{-m_\pi r}  + \dots\; \,,\\
	\label{Eq:p-large-r}
	p(r) 	& = -\,\frac16\;\frac{F_\pi^2\;R_0^4}{r^6}\;
		(6 + 12 m_\pi r + 13 m_\pi^2 r^2 + 10m_\pi^3 r^3 +4 m_\pi^4 r^4)
		\,e^{-m_\pi r} + \dots\;  \,,\\
	\label{Eq:s-large-r}
	s(r) 	& = \phantom{-}\,\frac12\;\frac{F_\pi^2\;R_0^4}{r^6}\;
		(6 + 12 m_\pi r+14 m_\pi^2 r^2+\;8 m_\pi^3 r^3\,+2 m_\pi^4 r^4)
		\,e^{-m_\pi r} + \dots\; \,.
\end{align}
\end{subequations}
In the chiral limit one finds the behavior quoted in 
Eqs.~(\ref{Eq:T00-large-r-0}--\ref{Eq:p-large-r-0}) in the main text, 
and for $m_\pi\neq 0$ one obtains large-distance behavior 
\begin{subequations}\label{Eq:EM-FF-large-r-mpi}
\begin{align}
	\label{Eq:T00-large-r-mpi}
	T_{00}(r)& =\phantom{-}\;
		F_\pi^2\;R_0^4\;\frac{m_\pi^4}{r^2}\;e^{-m_\pi r} + \dots\;\,,\\
	\label{Eq:p-large-r-mpi}
	p(r) 	& = -\frac23\;
		F_\pi^2\;R_0^4\;\frac{m_\pi^4}{r^2}\;e^{-m_\pi r} + \dots\;\,,\\
	\label{Eq:s-large-r-mpi}
	s(r) 	& = \phantom{-}\;
		F_\pi^2\;R_0^4\;\frac{m_\pi^4}{r^2}\;e^{-m_\pi r} + \dots\;\,,
\end{align}
\end{subequations}
Notice that Eq.~(\ref{Eq:R0-gA}) is valid only in chiral limit. 
For physical pion masses Eq.~(\ref{Eq:R0-gA}) approximates the respective 
model prediction for $g_A$ within $5\,\%$ \cite{Cebulla:2007ei}.

Although derived in soliton models, these results are practically 
model-independent. In particular, it was shown that they
imply the correct chiral behavior of the EMT form factors which
coincides with chiral perturbation theory 
\cite{Chen:2001pv,Diehl:2006ya} if one considers that
the large-$N_c$ limit and chiral limit do not commute
\cite{Goeke:2007fp}.
The non-commutativity of these limits is caused by the special
role of the $\Delta$-resonance. In the large-$N_c$ limit the
$\Delta$-nucleon mass splitting vanishes,
\be\label{Eq:Delta-N-mass-splitting}
	M_\Delta-M_N \sim {\cal O}(N_c^{-1}),
\ee
such that chiral loops with the $\Delta$-resonance as intermediate state 
contribute on the same footing as nucleon intermediate states to chiral 
properties. The contribution of the $\Delta$ to scalar-isoscalar quantities
in the large-$N_c$ limit is exactly two times larger than that of the nucleon 
\cite{Cohen:1992uy}. Therefore e.g.\ the leading non-analytic contributions
to the $D$-term derived from soliton models are 3 times larger than in
chiral perturbation theory \cite{Goeke:2007fp}. 
We have taken this into account in Sec.~\ref{Sec-3:condition-for-bound-state}
in our estimates of the Calogero bounds for $\alpha$ by reducing the
coefficient in the large-$r$ asymptotics of $V_{\rm eff}(r)$ by factor 3.
This resulting bound is lower and more realistic bound for $N_c=3$ colors.

It is interesting to inspect the local criterion (\ref{Eq:local-criterion-1}) 
at asymptotic distances. 
In the chiral limit the compliance of with (\ref{Eq:local-criterion-1}) 
is evident. But for finite $m_\pi$ the leading terms from 
(\ref{Eq:EM-FF-large-r}), i.e.\ the terms displayed in 
Eq.~(\ref{Eq:T00-large-r-mpi}) in the main text, cancel out exactly
and the criterion (\ref{Eq:local-criterion-1}) is fulfilled by the
subleading chiral terms. The results for both cases are
\begin{align}\begin{subequations}
	\frac23\,s(r) + p(r) = F_\pi^2\;R_0^4\;\times
	\begin{cases}
	\displaystyle\frac{1}{r^6} & \mbox{for $m_\pi=0$}, \\
	\displaystyle\frac{m_\pi^3}{r^3}\;e^{-m_\pi r} & \mbox{for $m_\pi\neq0$}.
\end{cases}
\end{subequations}\end{align}


\end{document}